\documentclass[preprint,notoc]{JHEP3}
\usepackage{epsfig}
\def\mtmin{m_T^{\rm min}}

\def\eslt{E_T^{\rm miss}}

\def\to{\rightarrow}

\def\bi{\begin{itemize}}
\def\ei{\end{itemize}}

\def\tb{\tilde b}

\def\tst{\tilde t}

\def\tg{\tilde g}

\def\tell{\tilde\ell}
\def\tq{\tilde q}

\def\tw{\widetilde W}
\def\tz{\widetilde Z}
\def\delew{\Delta_{\rm EW}}
\def\delbg{\Delta_{\rm BG}}
\def\delhs{\Delta_{\rm HS}}
\def\be{\begin{equation}}  
\def\ee{\end{equation}}  
\def\bea{\begin{eqnarray}}  
\def\eea{\end{eqnarray}}  
\def\beas{\begin{eqnarray*}}  
\def\eeas{\end{eqnarray*}}

\newcommand\prD[3]{{\it Phys.\ Rev.\ }{\bf D #1} (#2) #3}

\title{Radiatively-driven natural supersymmetry at the LHC
}
\author{Howard Baer$^a$, Vernon Barger$^b$, Peisi Huang$^{c,d}$, 
Dan Mickelson$^a$, Azar Mustafayev$^e$, Warintorn Sreethawong$^f$ and 
Xerxes Tata$^e$\\
$^a$Dept. of Physics and Astronomy, University of Oklahoma, Norman, OK 73019, USA\\
$^b$Dept. of Physics, University of Wisconsin, Madison, WI 53706, USA\\
$^c$Enrico Fermi Institute, University of Chicago, Chicago, IL 60637\\
$^d$High Energy Physics Division, Argonne National Laboratory, Argonne, IL 60439\\
$^e$Dept. of Physics and Astronomy, University of Hawaii, Honolulu, HI 96822, USA\\
$^f$School of Physics, Suranaree University of Technology, Nakhon Ratchasima 30000, Thailand\\
E-mail: \email{baer@nhn.ou.edu}, \email{barger@pheno.wisc.edu}, \email{peisi@uchicago.edu},
\email{mickelso@nhn.ou.edu}, \email{azar@phys.hawaii.edu}, \email{wsreethawong@hotmail.com}, 
\email{tata@phys.hawaii.edu}
}

\preprint{\vbox{UH-511-1218-13}}

\abstract{Radiatively-driven natural supersymmetry (RNS) potentially
reconciles the $Z$ and Higgs boson masses close to $\sim 100$~GeV with
gluinos and squarks lying beyond the TeV scale.  Requiring no large
cancellations at the electroweak scale in constructing $M_Z=91.2$~GeV
while maintaining a light Higgs scalar with $m_h\simeq 125$~GeV implies
a sparticle mass spectrum including light higgsinos with mass $\sim
100-300$~GeV, electroweak gauginos in the $300-1200$~GeV range, gluinos
at $1-4$~TeV and top/bottom squarks in the 1-4~TeV range (probably
beyond LHC reach), while first/second generation matter scalars can
exist in the 5-30~TeV range (far beyond LHC reach).  We investigate
several characteristic signals for RNS at LHC14.  Gluino pair production
yields a reach up to $m_{\tg}\sim 1.7$~TeV for 300~fb$^{-1}$.  Wino pair
production -- $pp\to\tw_2\tz_4 \ {\rm and}\ \tw_2\tw_2$ -- leads to a
unique same-sign diboson (SSdB) signature accompanied by modest jet
activity from daughter higgsino decays; this signature provides the best
reach up to $m_{\tg}\sim 2.1$~TeV within this framework.  Wino pair
production also leads to final states with $(WZ\to 3\ell )+\eslt$ as
well as $4\ell+\eslt$ which give confirmatory signals up to $m_{\tg}\sim
1.4$~TeV.  Directly produced light higgsinos yield a clean, soft
trilepton signature (due to very low visible energy release) which can
be visible, but only for a not-too-small a $\tz_2-\tz_1$ mass gap.  The
clean SSdB signal -- as well as the distinctive mass shape of the
dilepton mass distribution from $\tz_{2,3}\to\tz_1\ell\ell$ decays if
this is accessible -- will mark the presence of light higgsinos which
are necessary for natural SUSY.  While an $e^+e^-$ collider operating
with $\sqrt{s} \sim 600$~GeV should unequivocally reveal the predicted
light higgsinos, the RNS model with $m_{1/2}\gtrsim 1$~TeV may
elude all LHC14 search strategies even while maintaining a high degree
of electroweak naturalness.} 

\begin{document}

\section{Introduction to radiative natural SUSY}
\label{sec:intro}

The CERN LHC has gathered around 5~fb$^{-1}$ of data at $\sqrt{s}=7$~TeV
and over 20~fb$^{-1}$ at $\sqrt{s}=8$~TeV.  While this data set
was sufficient for the spectacular discovery of the Higgs boson at $m_h
=125.5\pm 0.5$~GeV (ATLAS/CMS combined)\cite{atlas_h,cms_h}, so far
there is no evidence for the production of supersymmetric (SUSY) matter.  The
lack of evidence for SUSY has already lead to new strong mass
limits\cite{atlas_susy,cms_susy} such as 
\bi
\item $m_{\tg}\gtrsim 1.7$~TeV for $m_{\tq}\simeq m_{\tg}$ and 
\item $m_{\tg}\gtrsim 1.1-1.3$~TeV for $m_{\tq}\gg m_{\tg}$, depending
  on the final state topology, 
\ei 
in the context
of the popular mSUGRA/CMSSM model\cite{msugra}.  In addition, the rather
large value of $m_h\simeq 125$~GeV requires highly mixed top squarks
with mass typically beyond the TeV scale \cite{h125}.  These TeV-scale
mass limits have led many physicists to question whether weak-scale supersymmetry {\it naturally} accommodates the weak scale, as typified by the
values of the $W^\pm$, $Z$ and $h$ masses $\sim 100$~GeV.

Specifically, these limits seemingly require increased cancellations
between the various terms in the well-known expression for
the $Z$-boson mass obtained from the minimization of the one loop MSSM
Higgs potential,
\be 
\frac{M_Z^2}{2} = \frac{m_{H_d}^2+\Sigma_d^d - (m_{H_u}^2+\Sigma_u^u) \tan^2\beta}{\tan^2\beta -1} -\mu^2 ,
\label{eq:mZs}
\ee 
where the $\Sigma$s contain 1-loop corrections to the Higgs field potential 
(expressions are given in the Appendix of Ref.~\cite{rns}). 
These show that {\it e.g.} the $\Sigma_u^u(\tst_{1,2})$ grow rapidly with the 
top-squark mass\footnote{The gluino can
also not be too heavy since renormalization effects from a heavy gluino
tend to make top squarks heavy in models defined at very high energy
scales.} unless model parameters are correlated in a special way as discussed below. 
Here we will require no large cancellations between the various
contributions to $M_Z^2$, meaning each term on the right-hand-side of
Eq.~(\ref{eq:mZs}) should have a magnitude comparable to $M_Z^2/2$.  Noting that all
entries in Eq.~(\ref{eq:mZs}) are defined at the weak scale, we were motivated
to define the electroweak (EW) fine-tuning parameter
\be 
\delew \equiv max_i \left|C_i\right|/(M_Z^2/2) .
\ee 
Here, $C_{H_d}=m_{H_d}^2/(\tan^2\beta -1)$, $C_{H_u}=-m_{H_u}^2\tan^2\beta /(\tan^2\beta -1)$ and $C_\mu =-\mu^2$. 
Also, $C_{\Sigma_u^u(k)} =-\Sigma_u^u(k)\tan^2\beta /(\tan^2\beta -1)$ and $C_{\Sigma_d^d(k)}=\Sigma_d^d(k)/(\tan^2\beta -1)$, 
where $k$ labels the various loop contributions included in Eq.~(\ref{eq:mZs}).

We stress that because $\delew$ depends only on the weak scale, it {\it
does not include} information about any possible high scale (HS) origin
of SUSY masses and couplings, and the potentially large logarithms that
could enter the right-hand-side of (\ref{eq:mZs}) at the loop-level in
HS models such as mSUGRA. The effect of these logarithms is captured in
the usual fine-tuning measures \cite{kn,papucci,ah,ns,sugra,rns} such as
$\delhs$ or $\delbg$, where the latter takes into account any
correlations among parameters that may be present. In contrast, $\delew$
captures the {\it minimal} fine-tuning that must be present. The utility
of $\delew$ stems from the fact that $\delew$ is essentially determined
by the SUSY spectrum, independently of the HS dynamics that generates
the spectrum. SUSY models with spectra that lead to large values of
$\delew$ are necessarily fine-tuned, whereas a spectrum with low values
of $\delew$ leaves open the possibility of finding an underlying
framework where the fine-tuning is small\cite{sugra,rns,comp}.

In order to achieve low values of $\delew$, it is necessary that
$-m_{H_u}^2$, $\mu^2$ and $-\Sigma_u^u$ all be within a factor of a few  of 
$M_Z^2/2$ \cite{ltr,rns}. This requires, 
\bi
\item the higgsino mass $|\mu|$ to lie in the $100-300$~GeV range for
  $\delew \lesssim 30$,\footnote{The connection between fine-tuning and
  the higgsino mass breaks down if the dominant contribution to the
  higgsino mass is SUSY breaking \cite{sundrum}. If there are no
  singlets that couple to the higgsinos, such a contribution would be
  soft. In all HS models that we are aware of, the higgsino masses have
  a supersymmetric origin. }

\item a value of $\sqrt{m_{H_u}^2(M_{\rm GUT})}\sim (1.3-2)m_0$ 
so that $m_{H_u}^2$ is driven radiatively to small negative values at the
weak scale, leading to $m_{H_u}^2({\rm weak})\sim -M_Z^2/2$, and
\item large stop mixing from $A_0\sim \pm 1.6 m_0$, that soften top-squark radiative corrections 
whilst raising $m_h$ to the $\sim 125$~GeV level.
\ei

In the well-known mSUGRA/CMSSM model, the lowest value of $\delew$ which
is found after respecting sparticle and Higgs mass constraints is $\sim
200$: thus, this model would necessarily be regarded as
fine-tuned\cite{sugra}.  In the two-extra-parameter non-universal Higgs
model\cite{nuhm2} (NUHM2), $\delew$ as low as $7-10$ can be found when
$m_{H_u}^2({\rm GUT})$, $\mu$ and $A_0$ are as stated above.  In this
case, the NUHM2 model may be considered as an effective theory valid up
to scales $\Lambda =M_{GUT}$ which may contain a more constrained
theory, with fewer free parameters, whose correlated soft terms would
yield lower $\Delta_{BG}$ values than those found in the more general
effective theory.  The phenomenology of this {\it meta-theory} -- since
it is essentially determined by the weak-scale SUSY spectrum -- would be
much the same as that from the NUHM2 model with parameter choices which
yield low values of $\delew$.  A survey of the LHC collider signatures
which arise from the NUHM2 model -- with parameter choices which yield
low values of $\delew$ -- is the subject of this paper.

Detailed scans over NUHM2 parameter space in Ref.~\cite{rns} find that low EWFT of order
$\delew^{-1}\sim 3-15\%$ can be achieved while at the same time requiring LHC sparticle mass bounds 
and $m_h\sim 123-127$~GeV for the following parameter choices:
\bi
\item $m_0\sim 1-7$~TeV,
\item $m_{1/2}\sim 0.3-1.5$~TeV,
\item $A_0\sim \pm (1-2)m_0$,
\item $\tan\beta\sim 5-50$,
\item $\mu\sim 100-300$~GeV, 
\ei 
while $m_A$ may vary over a wide range
of values.  This framework has been dubbed {\it Radiatively-driven Natural
Supersymmetry} (RNS), since the required value of $-m_{H_u}^2(weak)\sim
M_Z^2$ is generated radiatively via running from the GUT scale.  The
sparticle mass spectrum of RNS differs sharply in the top-squark 
sector from what has been referred to as natural supersymmetry (NS) in the literature. 
For phenomenologically viable RNS scenarios, we find:
\bi
\item the presence of four higgsino-like states with mass $m_{\tw_1},\
m_{\tz_{1,2}}\sim 100-300$~GeV and with mass gap
$m_{\tz_2}-m_{\tz_1}\sim 10-30$~GeV,
\item $m_{\tst_1}\sim 1-2$~TeV, $m_{\tst_2,\tb_1}\sim 2-4$~TeV,
\item $m_{\tg}\sim 1-5$~TeV,
\item $m_{\tq,\tell}\sim 5-10$~TeV (first/second generation sfermions).
\ei
Much higher values of first/second generation scalar masses $m_{\tq,\tell}\sim 10-30$~TeV are allowed if non-universal
generations are implemented, {\it i.e.} $m_0(3)< m_0(1,2)$. 
We remark that such heavy first/second generation scalars allow for at least a partial
{\it decoupling solution} to the SUSY flavour and $CP$ problems. 
The RNS scenario retains the characteristic light higgsinos of previously studied NS models.
But for RNS, the top and bottom squark masses are considerably heavier than 
previous NS estimates, and are probably beyond LHC reach. 
Also, heavier values of $m_{\tg}$ reaching up to $\sim 5$~TeV are allowed in RNS 
as compared to previous NS models.

Our goal in this paper is to assess the prospects for CERN LHC operating
at $\sqrt{s}=14$~TeV and $30-300$~fb$^{-1}$ (or even higher) to discover
supersymmetry within the RNS context.  Previous work along this line has
already been presented in Ref.~\cite{lhcltr} where it was noted that
SUSY models with light higgsinos give rise to a qualitatively new
signature: same-sign diboson (SSdB) production accompanied by minimal
hadronic activity.  This signal arises from wino pair production: $pp\to
\tw_2^\pm\tz_4\to (W^\pm\tz_{1,2})(W^\pm\tw_1^\mp)$, where the higgsinos
$\tz_2$ and $\tw_1$ decay into very soft hadrons or leptons owing to
their small mass gaps with the LSP, which is assumed to be the lightest higgsino $\tz_1$. 
In Ref.~\cite{lhcltr}, it was shown that for higher luminosity values, the
SSdB signature yields a better reach for SUSY than gluino pair searches
(assuming unified gaugino masses), due to the rapidly falling $\tg\tg$
production cross section as compared to $\tw_2\tz_4$ production. 
In this paper, we provide a detailed treatment of a variety  of different signatures
expected at LHC14 for the RNS model.
 
Towards this end, in Sec.~\ref{sec:mline}, we construct a RNS model line
which contains all the generic features of RNS models, but with a
variable gluino mass.  In Sec.~\ref{sec:prod}, we show sparticle
production cross sections and branching fractions along the RNS model
line.  Sec.~\ref{sec:gg} examines prospects for discovering gluino pair
production via signals from their cascade decays.  If a signal is found,
then the shape of the mass distribution of opposite sign, same flavour
dileptons from $\tz_2\to\tz_1\ell\bar{\ell}$ decays of neutralinos
produced via cascade decays (or directly, see Sec.~\ref{sec:3l}),
characterizes models with light higgsinos, as emphasized in
Ref.~\cite{kn,heaya,kadala}.  In Sec.~\ref{sec:SSdB}, we examine aspects
of the characteristic same-sign diboson signature from SUSY models with
light higgsinos, previously presented in Ref.~\cite{lhcltr}.  In
Sec.~\ref{sec:WZ}, we examine prospects for LHC to detect the clean
trilepton signal arising from wino pair production followed by decay to
$WZ+\eslt$.  In Sec.~\ref{sec:4l}, we examine a novel $4\ell+\eslt$
signal from wino pair production.  Sec.~\ref{sec:3l} examines the
possibility of detecting directly produced higgsinos -- whose decays have
a very low energy release in the RNS framework -- in the soft trilepton
channel with low jet activity.  We end in Sec.~\ref{sec:conclude} with a
summary of our results along with a grand overview plot.

The main findings of our work is that LHC14 with 300~fb$^{-1}$ has the capability
to discover RNS SUSY models for $m_{\tg}$ values up to $\sim 2$~TeV in
the SSdB and $\tg\tg$ channels.  Since RNS models maintain low EW
naturalness for $m_{\tg}$ ranging up to 5~TeV, there remains a rather
large chunk of parameter space beyond the reach of LHC14. 
Direct higgsino detection at next generation WIMP direct detection experiments seemingly
probes the entire parameter space.  However, it will likely require an
$e^+e^-$ collider operating with $\sqrt{s}\sim 600$~GeV to either
discover the light higgsinos predicted by RNS models with $\delew ^{-1} < 3\%$, 
or to decisively rule these out.

\section{A radiative natural SUSY model line}
\label{sec:mline}

The radiative natural SUSY model automatically maintains the SUSY
success stories of gauge coupling unification and radiative breaking of
electroweak symmetry due to a large top quark mass.  These features
require the MSSM (possibly augmented by gauge singlets or additional
GUT multiplets) as the effective field theory up to a scale $\Lambda$,
which we take to be $M_{GUT}\simeq 2\times 10^{16}$~GeV.
The low value of $m_{H_u}^2({\rm weak})$ 
that is required to obtain small $\delew$ can always be realized 
via RG running, once the GUT-scale value of $m_{H_u}^2$ is decoupled 
from matter scalar masses.
In order to implement a low value of $|\mu |\in (100-300)$~GeV, we use
the 2-parameter non-universal Higgs model (NUHM2)\cite{nuhm2}, wherein
weak-scale values of $\mu$ and $m_A$ may be used as inputs in lieu of
GUT-scale values of $m_{H_u}^2$ and $m_{H_d}^2$.  Thus, we will adopt
the NUHM2 parameter set
\be
m_0,\ m_{1/2},\ A_0,\ \tan\beta,\ \mu,\ m_A \; ,
\ee
and also take $m_t=173.2$~GeV.
For our calculations, we adopt the Isajet 7.83~\cite{isajet} SUSY
spectrum generator Isasugra\cite{isasugra}.\footnote{Isasugra begins the
calculation of the sparticle mass spectrum with input $\overline{DR}$
gauge couplings and $f_b$, $f_\tau$ Yukawa couplings at the scale
$Q=M_Z$ ($f_t$ running begins at $Q=m_t$) and evolves the 6 couplings up
in energy to scale $Q=M_{GUT}$ (defined as the value $Q$ where
$g_1=g_2$) using two-loop RGEs.  At $Q=M_{GUT}$, the SSB boundary
conditions are input, and the set of 26 coupled two-loop MSSM
RGEs~\cite{mv} are evolved back down in scale to $Q=M_Z$.  Full two-loop
MSSM RGEs are used for soft term evolution, while the gauge and Yukawa
coupling evolution includes threshold effects in the one-loop
beta-functions, so the gauge and Yukawa couplings transition
continuously from the MSSM to SM effective theories as different mass
thresholds are passed.  In Isasugra, the values of SSB terms which mix
are frozen out at the scale $Q\equiv M_{\rm SUSY}=\sqrt{m_{\tst_1}
m_{\tst_2}}$, while non-mixing SSB terms are frozen out at their own
mass scale~\cite{isasugra}.  The scalar potential is minimized using the
RG-improved one-loop MSSM effective potential evaluated at an optimized
scale $Q=M_{SUSY}$ which accounts for leading two-loop
effects~\cite{haber}.  Once the tree-level sparticle mass spectrum is
computed, full one-loop radiative corrections are calculated for all
sparticle and Higgs boson masses, including complete one-loop weak scale
threshold corrections for the top, bottom and tau masses at scale
$Q=M_{SUSY}$~\cite{pbmz}.}

NUHM2 model parameter values leading to low $\delew\sim 10$ (RNS solutions) 
were found in Ref.~\cite{rns}. We use those results to
construct a RNS model line which features a variable gluino mass, via
\bea
m_0 &=&5\ {\rm TeV}, \nonumber \\
m_{1/2}&:&\ {\rm variable\ between\ 0.3-2\ TeV}, \nonumber \\
A_0&=&-1.6 m_0, \nonumber \\
\tan\beta &=& 15 ,  \\
\mu &=& 150\ {\rm GeV}, \nonumber \\
m_A &=& 1\ {\rm TeV} \nonumber .
\label{eq:param}
\eea 
The variation in $m_{1/2}$ corresponds to variation in $m_{\tg}$
from about $0.9$~TeV to $\sim 5$~TeV.

In Fig.~\ref{fig:delew}, we show the value of $\delew$ along the
RNS model line.  We see that $\delew$ begins around 11 at
$m_{1/2}\sim 300$~GeV and increases only mildly with $m_{1/2}$, reaching
$\delew\sim 20$ for $m_{1/2}$ as high as 1000~GeV.  This
corresponds to EWFT of $\sim 9\%$ at the low end of $m_{1/2}$ and $\sim
5\%$ at around $m_{1/2}\sim 1$~TeV. 
\FIGURE[tbh]{
\includegraphics[width=12cm,clip]{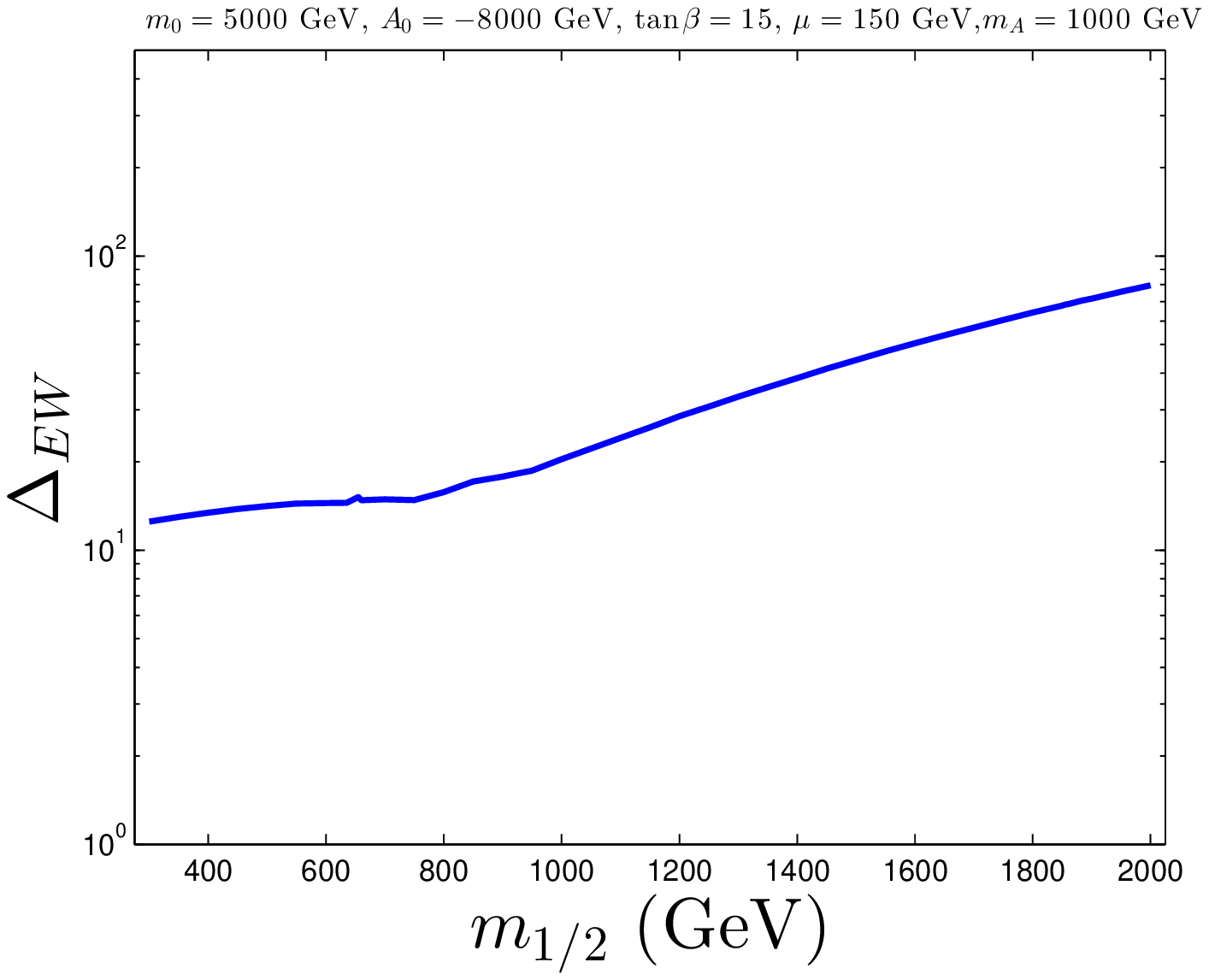}
\caption{Plot of $\delew$ versus $m_{1/2}$ along the RNS model line.
}
\label{fig:delew}}

In Fig.~\ref{fig:mass}, we plot various sparticle masses from the RNS
model line versus $m_{1/2}$. Along the model line, the value of $m_h$
varies from $124.4-125.2$~GeV, quite compatible with the recent
ATLAS/CMS Higgs resonance discovery\cite{atlas_h,cms_h}.  Also, since
$\mu$ is fixed at 150~GeV, we obtain a spectrum of higgsino-like
$\tw_1^\pm,\ \tz_1$ and $\tz_2$ states with mass $\sim
150$~GeV. However, along the model line, the mass gap
$m_{\tz_2}-m_{\tz_1}$ varies from 55.6~GeV for very low $m_{1/2}$ to just
under $\sim 10$~GeV if $m_{1/2}$ nears the values allowed
by $\delew\lesssim 30$, as shown in Fig.~\ref{fig:gap}.  
The behaviour of light chargino/neutralino masses
is easily understood since for low $m_{1/2}$ the weak scale gaugino mass
$M_1\simeq 136$~GeV and so the $\tz_1$ state is really a bino-higgsino
admixture, while at $m_{1/2}\sim 1$ TeV then $M_1\simeq 444$~GeV so that
$\tz_1$ is more nearly a pure higgsino state.
\FIGURE[tbh]{
\includegraphics[width=14cm,clip]{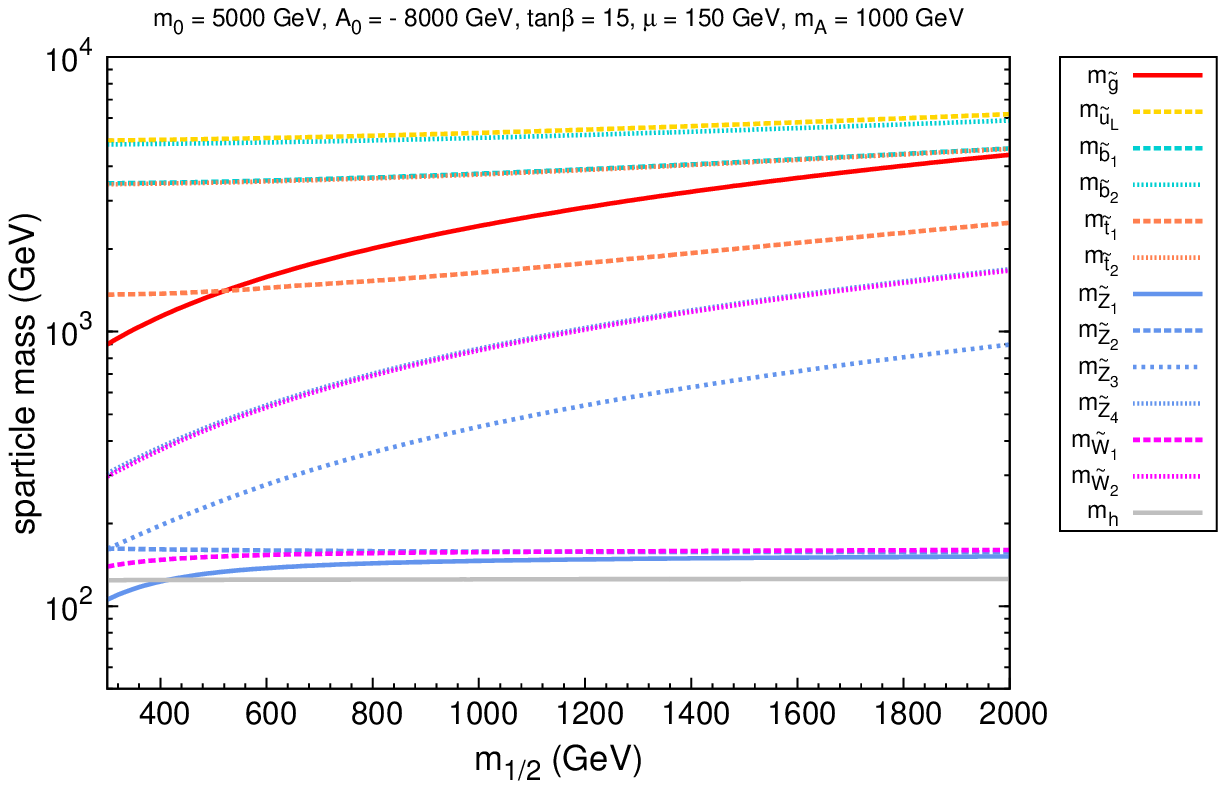}
\caption{Various sparticle masses versus $m_{1/2}$ for the RNS
model line
}
\label{fig:mass}}

After the higgsinos, the next lightest sparticles are the bino-like
$\tz_3$ -- whose mass varies between $160-900$~GeV -- and the wino-like $\tw_2^\pm$
and $\tz_4$ states -- whose masses vary between $300-1700$~GeV -- for the range of $m_{1/2}$ shown in the figure.
The solid red curve denoting $m_{\tg}$ varies between $900-4500$~GeV.
The red-dashed $m_{\tst_1}$ contour varies between
$1360-2500$~GeV over the $m_{1/2}$ range shown in the figure; the line crosses the
$m_{\tg}$ curve at $m_{1/2}\sim 520$~GeV.  The first/second generation
squarks and sleptons inhabit the multi-TeV range, and are far beyond the
reach of LHC14.
\FIGURE[tbh]{
\includegraphics[width=13cm,clip]{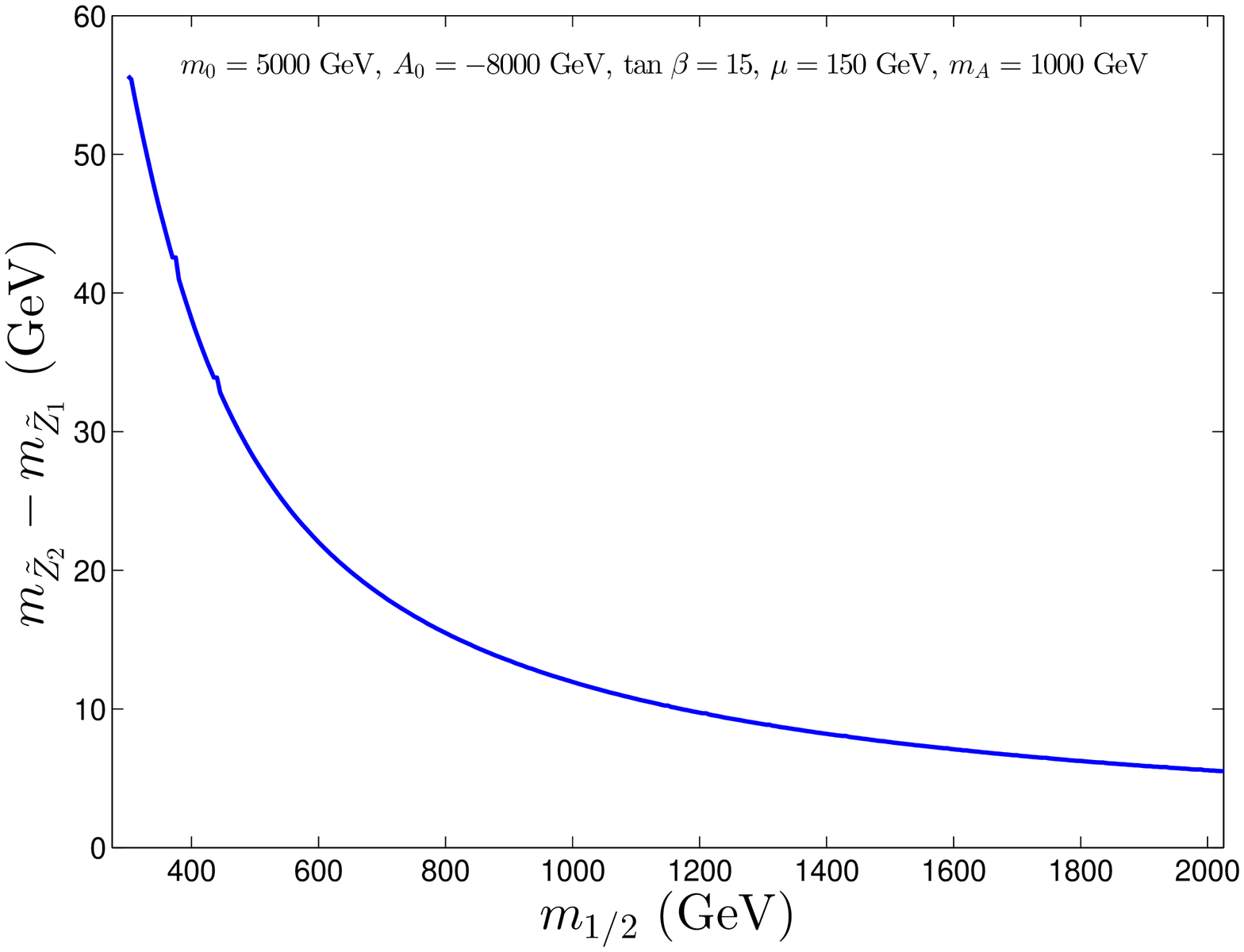}
\caption{The $m_{\tz_2}-m_{\tz_1}$ mass gap versus $m_{1/2}$ along the RNS
model line.
}
\label{fig:gap}}
%

\section{Sparticle production and decay in RNS}
\label{sec:prod}

In this Section, we show various sparticle pair production cross sections and
selected sparticle branching fractions along the RNS model line in order
to survey the panorama of the LHC detection possibilities.

\subsection{Sparticle production at LHC}

In Fig.~\ref{fig:xsec}, we show various sparticle pair production cross sections
at LHC for {\it a}) $\sqrt{s}=8$~TeV and {\it b}) $\sqrt{s}=14$~TeV versus $m_{1/2}$
along the RNS model line. 
We use Prospino\cite{prospino} to generate the cross sections at NLO in QCD.
\FIGURE[tbh]{
\includegraphics[width=12cm,clip]{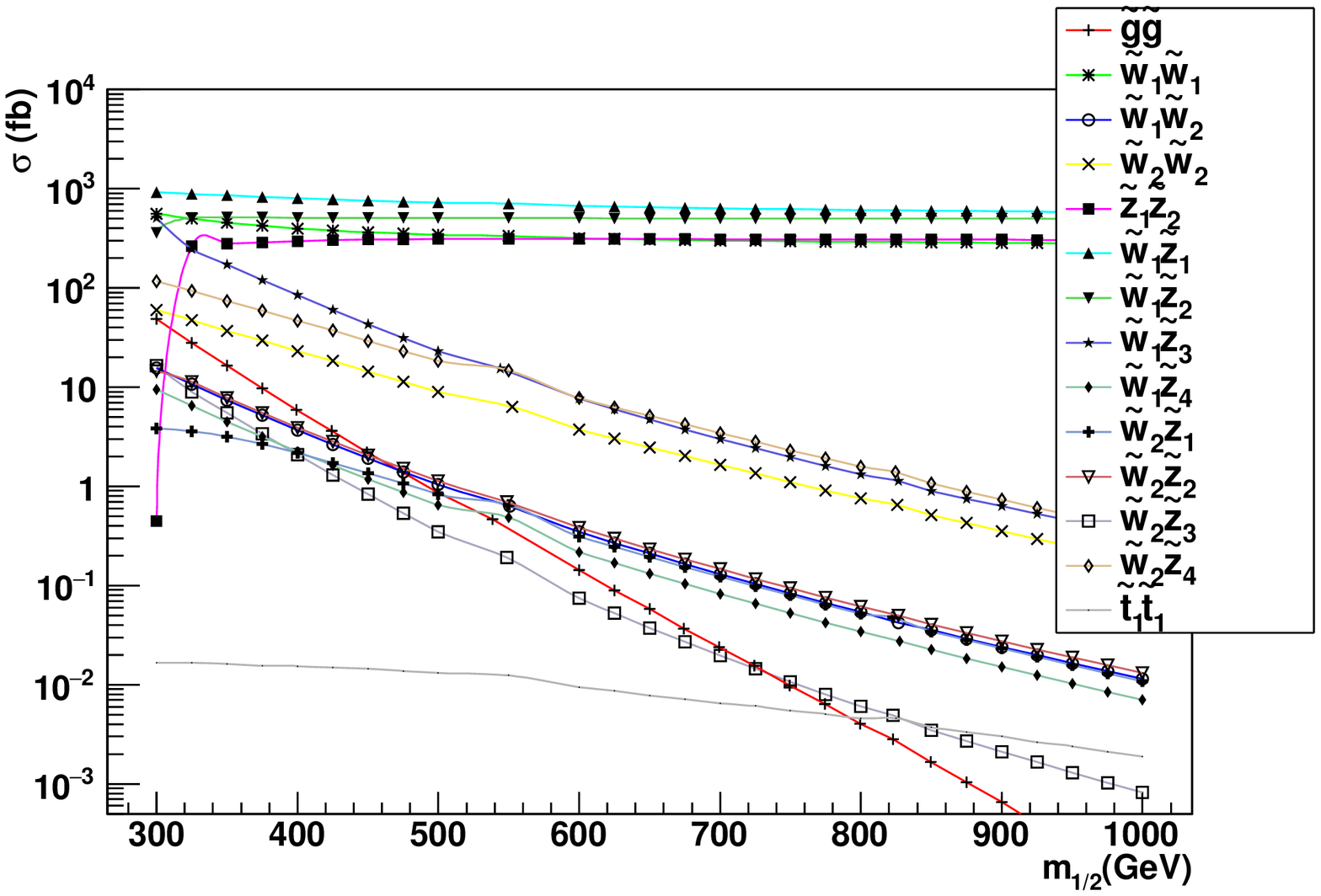}
\includegraphics[width=12cm,clip]{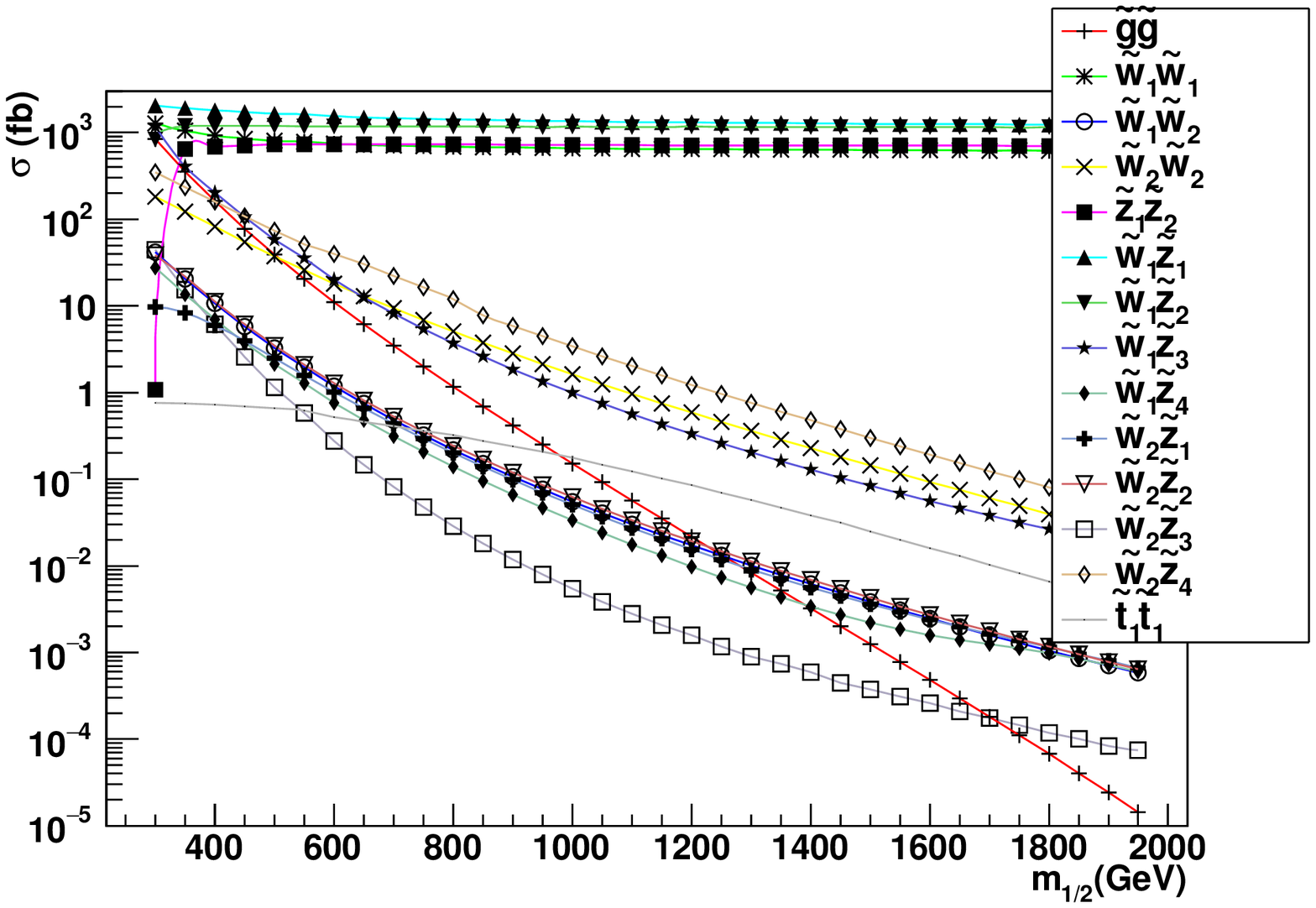}
\caption{Plot of various NLO sparticle pair production cross sections 
versus $m_{1/2}$ along the RNS model line for $pp$ collisions at
{\it a}) $\sqrt{s}=8$~TeV and {\it b}) $\sqrt{s}=14$~TeV.
}
\label{fig:xsec}}

From Fig.~\ref{fig:xsec}, we see that the four higgsino pair production
reactions -- $pp\to \tw_1^\pm\tz_1$, $\tw_1^\pm\tz_2$, $\tw_1^+\tw_1^-$
and $\tz_1\tz_2$ -- all occur at comparable rates of $\sim 500$~fb at
LHC8 and of $\sim 1000$~fb at LHC14. These cross sections
are nearly flat with increasing $m_{1/2}$ since they mainly depend on
$\mu$ which is fixed at 150~GeV along the model line.

The gluino pair production cross section -- $\sigma (pp\to\tg\tg X)$ -- is
denoted by the red curve with pluses. (As a guide, we note that
$m_{\tg}\simeq 2.6 m_{1/2}$.)  While the $\tg\tg$ production cross
section is large at $m_{1/2}\sim 300$~GeV (corresponding to $m_{\tg}\sim 900$~GeV), it
drops off rapidly with increasing values of $m_{1/2}$: it is likely to be inconsequential 
for even LHC14 searches for the upper range of $m_{1/2}\gtrsim 1$~TeV unless extremely 
high integrated luminosities are attained. 

Also of importance are the gaugino pair production reactions: wino pair
production $pp\to\tw_2^\pm\tz_4$ and $\tw_2^+\tw_2^-$, and also
$\tw_1^\pm\tz_3$ which proceeds via the higgsino component of
the bino-like $\tz_3$.  Wino pair production can be large due to the
large $SU(2)$ triplet gauge coupling. The cross section for this drops
off much less sharply than that for $\tg\tg$ production since the
wino masses are much smaller than the gluino mass. The cross section for
$\tw_1\tz_3$ production falls off faster than the wino production 
cross section because the higgsino content of $\tz_3$ drops off with
increasing $m_{1/2}$. We will see below
that these reactions constitute the largest {\it observable} SUSY cross
sections over most of the range of $m_{1/2}$.

For comparison, we also show cross sections for the pair production of
top squarks, the lightest sfermions in RNS.  The tiny
$\tst_1\bar{\tst}_1$ production cross section at $\sqrt{s}=8$ TeV
precludes any possibility of stop detection at LHC8.  Since for RNS
models $m_{\tst_1}$ lies within the $1-2$~TeV range (and
$m_{\tst_2,\tb_1}$ lies within $2-4$ TeV) over the range where the
fine-tuning is better than 3\%, it is clear that their detectability at
even high luminosity upgrades of LHC14 will be very difficult, and
certainly not extend over the entire allowed mass range
\cite{epjc,atlasstop,cmsstop}.  The top/bottom squark mass range in
radiatively-driven natural SUSY differs sharply from earlier versions of
natural SUSY\cite{kn,papucci,ah} where third generation squarks with
$m_{\tst_{1,2},\tb_1} \lesssim 600$~GeV are to be expected.  Such light
third generation squarks have difficulty generating a large enough
radiative correction to allow for $m_h\sim 125$ GeV, they lead to
anomalous contributions to $b\to s\gamma$ decay, and they likely should
have already been seen in LHC8 top squark search analyses.

In Fig.~\ref{fig:xsecmu}, we show selected electroweak-ino cross sections
versus $\mu$ for $m_{1/2}=750$~GeV along the RNS model line for LHC14. Here we see
that $\tw_1\tz_2$, $\tz_1\tz_2$ and $\tw_1^+\tw_1^-$ production are all
comparable and as high as $\sim 5000$~fb at $\mu\sim 100$~GeV. They drop
to the vicinity of $\sim 10^2$~fb at $\mu\sim 300$~GeV. It is
conceivable that a monojet search for $pp\to\tz_1\tz_1$ production
including initial state radiation of a hard gluon could reveal
higgsino-like dark matter production at LHC\cite{c_han}.  
The cross section for this reaction (and also for $\tz_2\tz_2$ pair production) is several
orders of magnitude below the other cross sections because the coupling
of identical higgsino-like neutralinos
to $Z$ is dynamically suppressed, and so seems much
more challenging to detect. However, this hard monojet signal may receive significant
contributions from $\tw_1^\pm\tw_1^\mp$, $\tz_1\tz_2$ and $\tw_1^\pm\tz_2$ production
processes which have much larger cross sections since the daughters from
$\tw_1$ and $\tz_2$ decays are expected to be soft at least for higher $m_{1/2}$ values. 
\FIGURE[tbh]{
\includegraphics[width=12cm,clip]{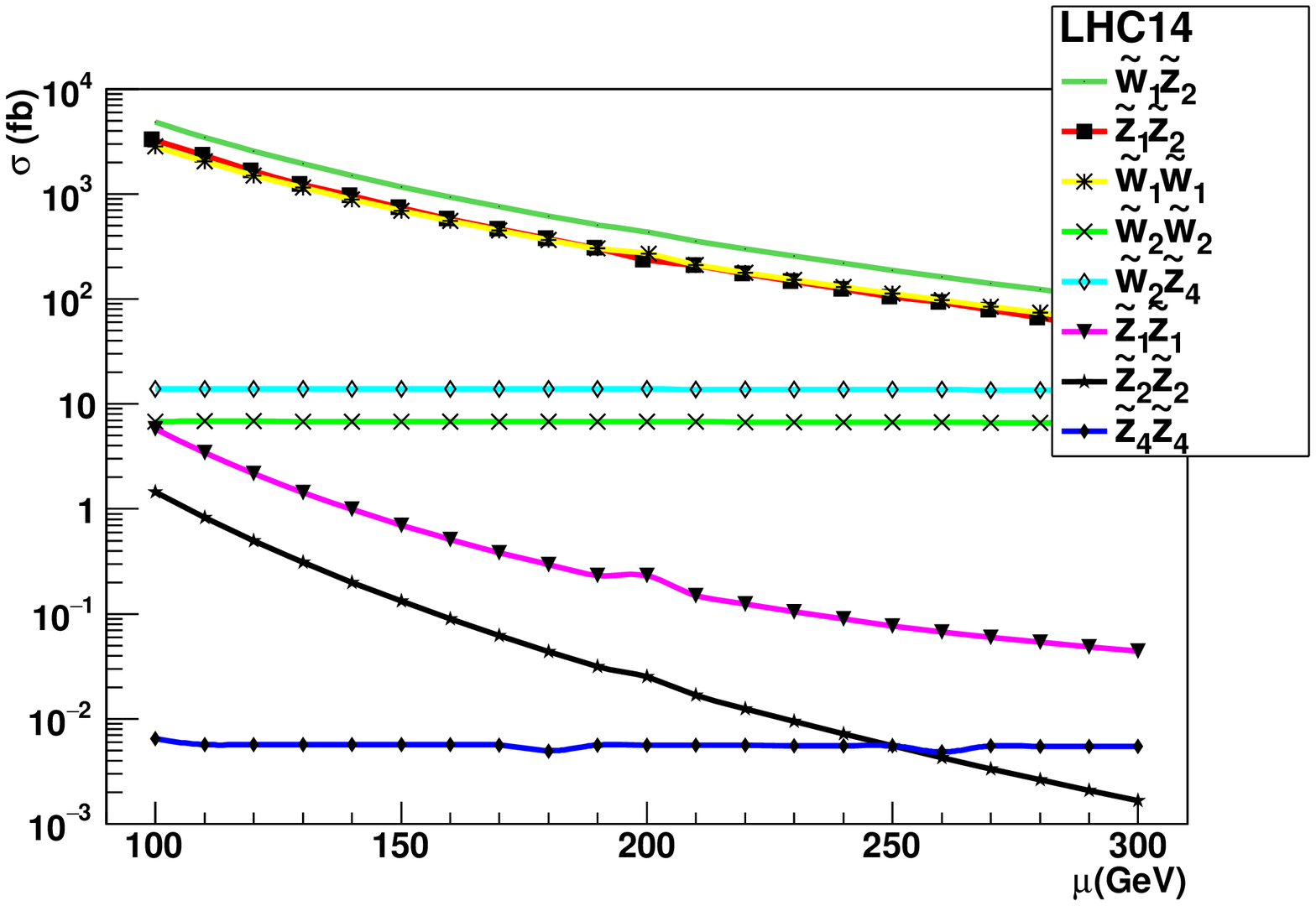}
\caption{Plot of various NLO electroweak-ino pair production cross sections 
versus $\mu$ for the RNS model line with $m_{1/2}=750$~GeV 
for $pp$ collisions at 14~TeV.
}
\label{fig:xsecmu}}

\subsection{Sparticle branching fractions}

In Fig.~\ref{fig:bf}, we show various sparticle branching fractions for
sparticles most accessible at the LHC, {\it i.e.}
{\it a}) $\tg$, {\it b}) $\tst_1$, {\it c}) $\tz_2$, {\it d}) $\tz_3$, {\it e}) $\tz_4$,
and {\it f}) $\tw_2$.
\FIGURE[tbh]{
\includegraphics[width=7cm,clip]{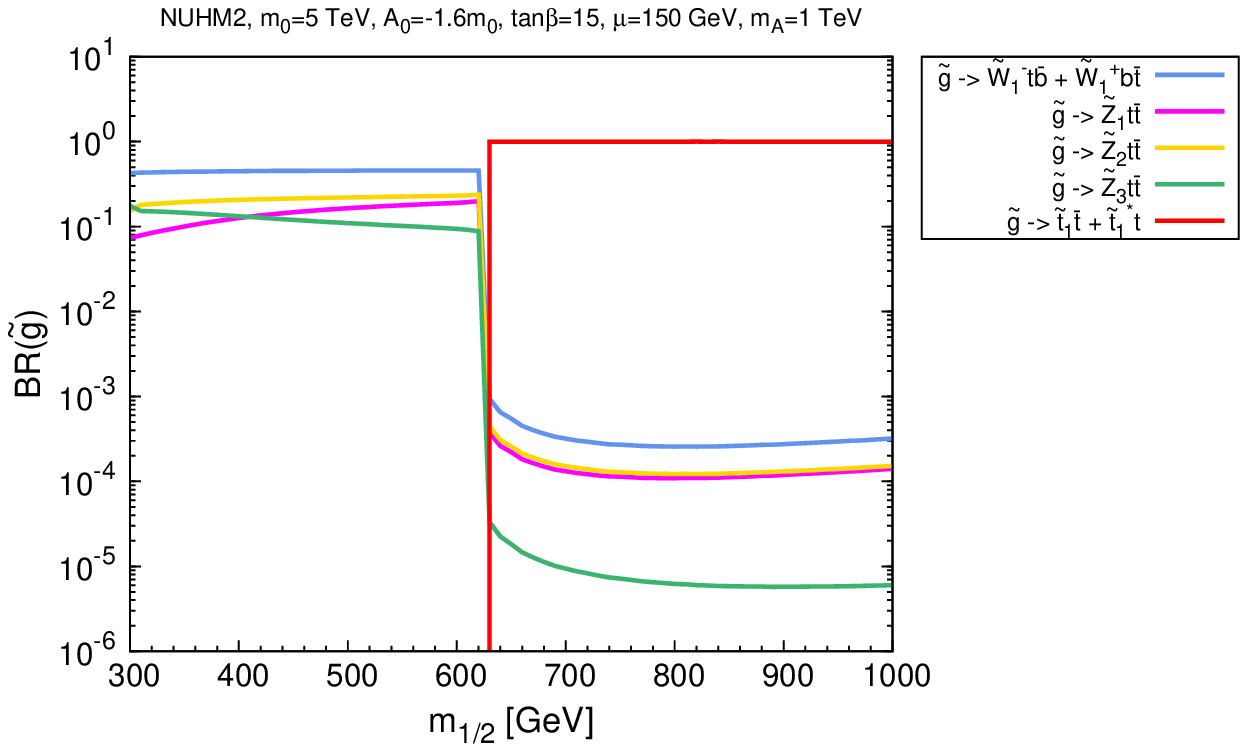}
\includegraphics[width=7cm,clip]{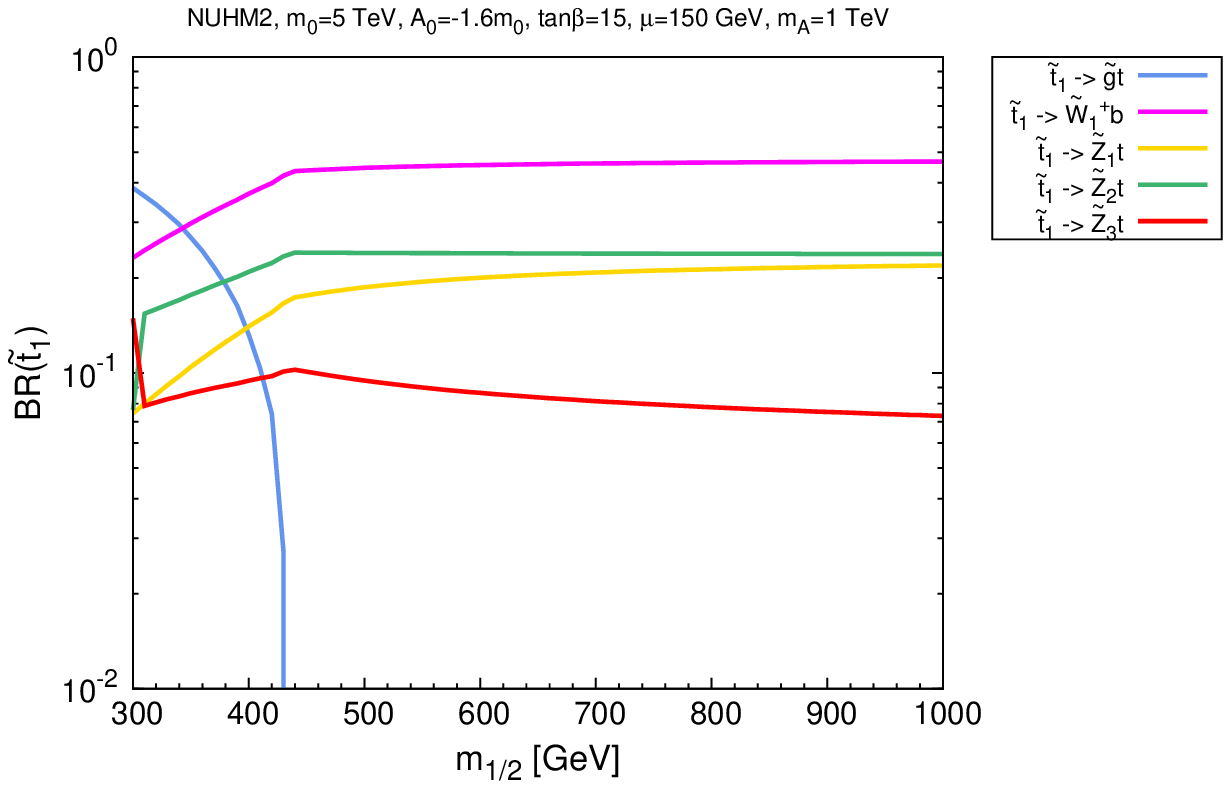}
\includegraphics[width=7cm,clip]{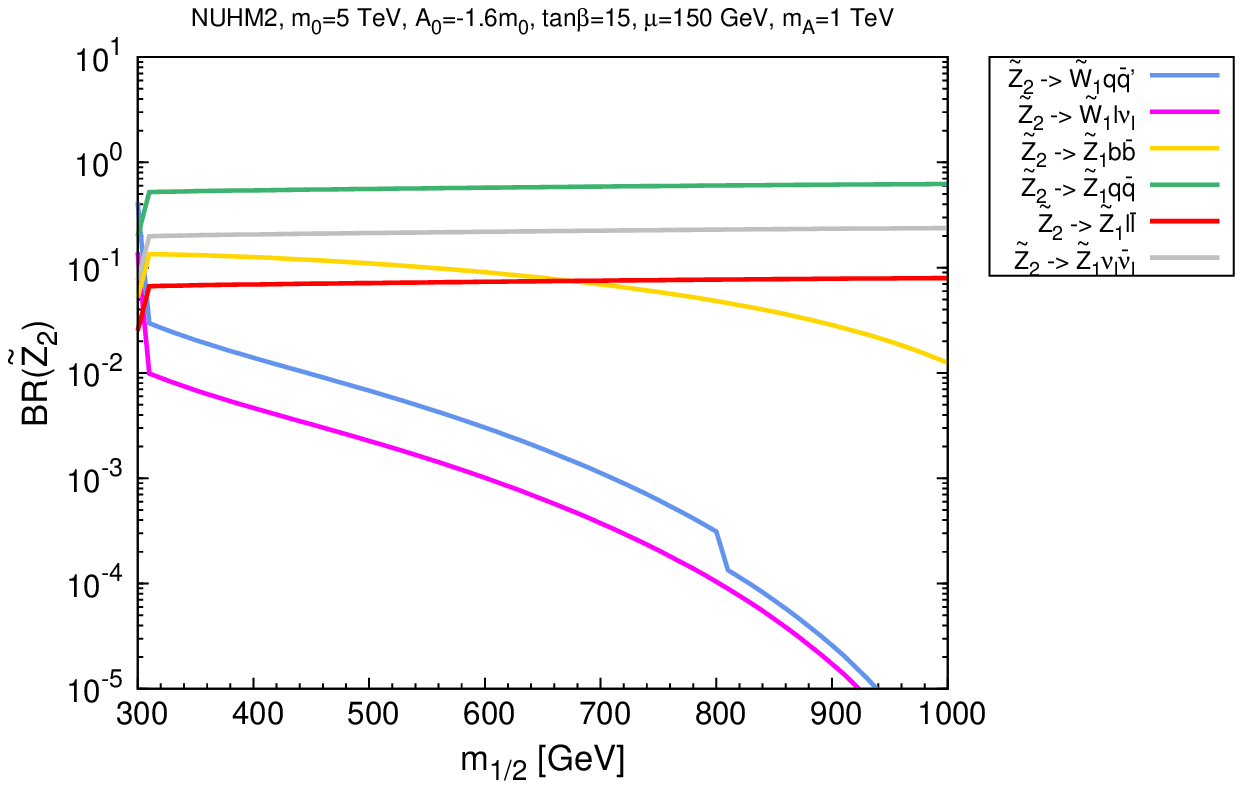}
\includegraphics[width=7cm,clip]{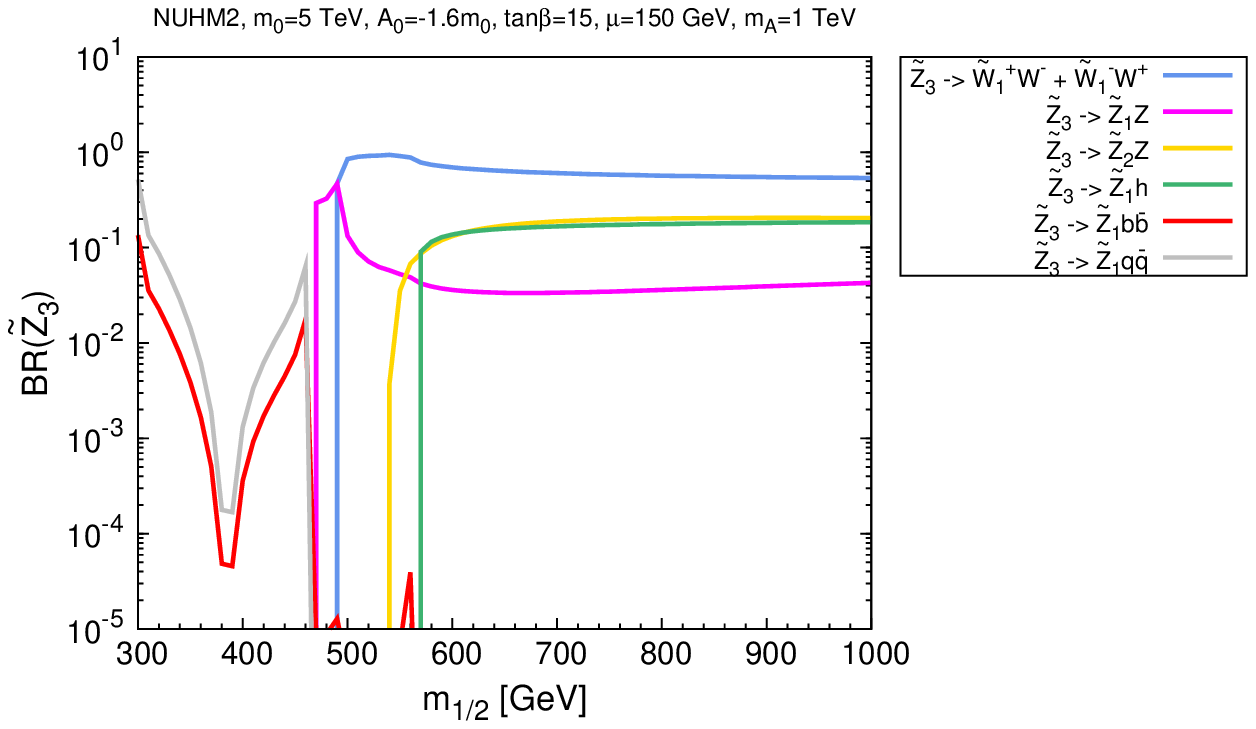}
\includegraphics[width=7cm,clip]{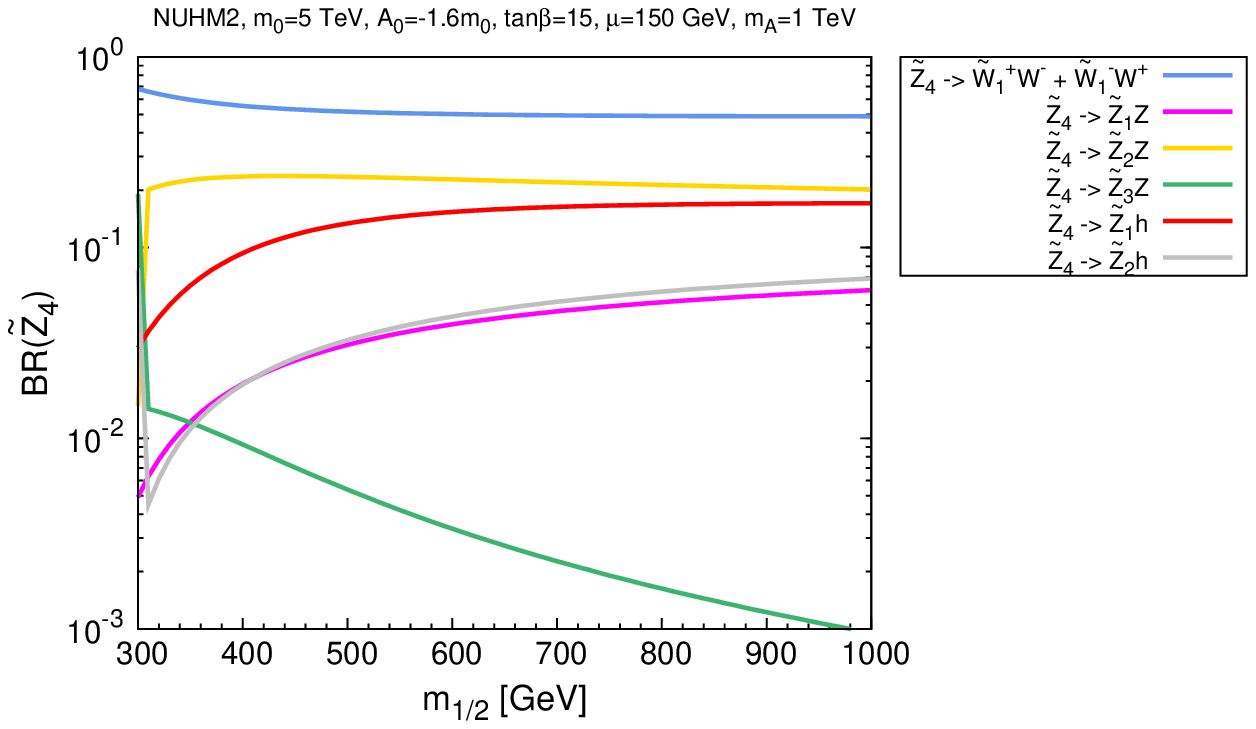}
\includegraphics[width=7cm,clip]{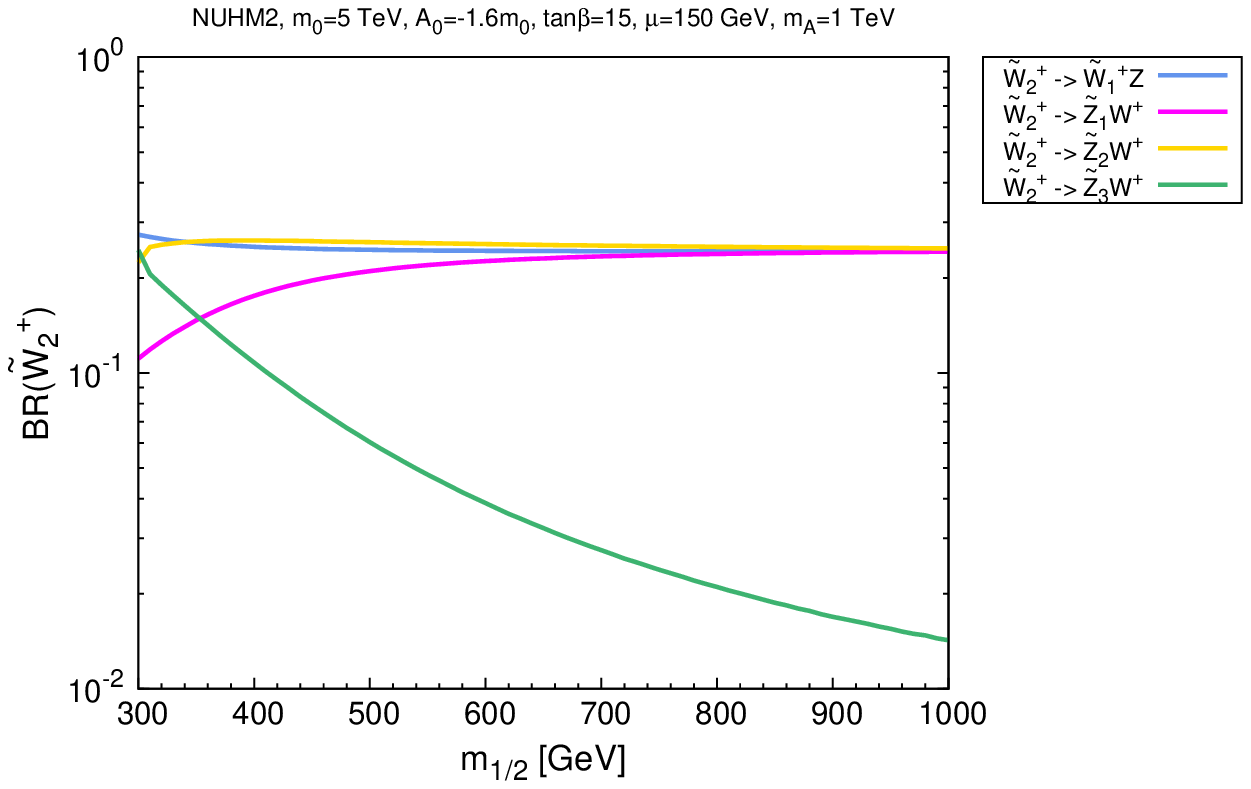}
\caption{Plot of various sparticle branching fractions versus $m_{1/2}$ along the RNS model line.
}
\label{fig:bf}}
From frame {\it a}), we see that for the lower portion of $m_{1/2}$
corresponding to $m_{\tg}\lesssim 1.8$~TeV, the gluino decays via 3-body
modes into $tb\tw_1$ and $t\bar{t}\tz_{1,2,3}$ states. For heavier
$m_{\tg}\gtrsim 1.8$~TeV, the 2-body modes $\tg\to t\tst_1$ open up
and dominate the decays. Thus, we expect the gluino pair production
events to be rich in $b$-jet activity\cite{btw1,btw2}. In the case where
$\tg\to t\tst_1$, it is important to know how $\tst_1$ decays. This is
shown in frame {\it b}). For the very lowest $m_{1/2}$ values, 
the $\tst_1\to t\tg$ decay mode is open and is dominant. However, as $m_{1/2}$
increases, this mode quickly closes and instead $\tst_1$ decays into
$b\tw_1$ or $t\tz_{1,2,3}$.

In frame {\it c}), we show the $\tz_2$ decay modes. Since the
$m_{\tz_2}-m_{\tz_1}$ mass gap ranges from $\sim 55$~GeV 
(already excluded for this model line by LHC8 gluino searches) 
to $\sim 10$~GeV along the model line, then $\tz_2$ always decays dominantly to
3-body modes $\to\tw_1 f\bar{f}'$ or $\to\tz_1 f\bar{f}$, where $f$
stands for kinematically accessible SM fermions. As mentioned earlier,
since the $\tz_2-\tz_1$ mass gap is small and the released decay energy
is shared between three particles, then the decay products from $\tz_2$ decay
are usually very soft -- in the few~GeV range. 
The light chargino (branching fractions not shown) decays into $\tz_1
f\bar{f}'$ mainly via $W^*$, where the $f$ and $\bar{f}'$ are again
typically rather soft.

In frame {\it d}) we show the bino-like $\tz_3$ decays. Here $\tz_3\to
\tw_1 f\bar{f}'$ or $\tz_{1,2}f\bar{f}$ for $m_{1/2}\lesssim
500$~GeV. For heavier $m_{\tz_3}\gtrsim 220$~GeV (this value, of course,
depends on our choice of $\mu$), the 2-body decays
$\tz_3\to \tw_1^\pm W^\mp$ and $\tz_{1,2}Z$ and $\tz_1 h$ turn on,
leading to production of vector bosons and Higgs bosons in the SUSY
events. 

In frames {\it e}) and {\it f}), we show the neutral $\tz_4$ and charged
$\tw_2^\pm$ wino branching fractions. We see that
$\tz_4\to \tw_1^\pm W^\mp$ mode dominates over the entire range of
$m_{1/2}$. The subdominant decay modes $\tz_4\to \tz_{1,2}Z$ and
$\tz_{1,2} h$ can also be important and occur at significant rates.  
The sizeable branching ratio for the decay $\tz_4 \to Z \tz_1$
may be surprising at first glance since $\tz_4$ is dominantly a wino
while $\tz_1$ is mostly a higgsino, so
that the $Z\tz_4\tz_1$ coupling should be suppressed by the small
higgsino content $\sim M_Z/M_2$ (assuming $M_2\gg |\mu|$) of $\tz_4$. For
heavy $\tz_4$, this suppression is compensated for by the fact that the
amplitude for decay to the {\em longitudinally polarized} $Z$ boson is
enhanced by $\sim |\mu|/M_Z$. As a result, for
$M_2 \gg M_Z, |\mu|$, the branching fractions for decays to $Z$ and to $h$
become comparable. This is discussed in
detail in Ref.~\cite{han}. In the
case of $\tw_2$ decay shown in frame {\it f}), we see that $\tw_2\to
\tw_1 Z$ or $\tz_{1,2} W$ or $\tw_1 h$ over the entire range of
$m_{1/2}$, leading again to production of gauge and Higgs bosons in
wino pair production events.  The dominant sparticle branching
fractions for $m_{1/2}=1$~TeV along the RNS model line are shown in
Table~\ref{tab:BF}.

\begin{table}
\begin{center}
\begin{tabular}{|l|r|r|}
\hline
Particle & dom. mode  &  BF \\
\hline
\hline
$\tg$    & $\tst_1 t$      &  $\sim 100\%$ \\
$\tst_1$ & $b\tw_1$        &  $\sim 50\%$ \\
$\tz_2$  & $\tz_1 f\bar{f}$ &  $\sim 100\%$ \\
$\tz_3$  & $\tw_1^\pm W^\mp$ &  $\sim 50\%$ \\
$\tz_4$  & $\tw_1^\pm W^\mp$ &  $\sim 50\%$ \\
$\tw_1$  & $\tz_1 f\bar{f}'$ &  $\sim 100\%$ \\
$\tw_2$  & $\tz_i W$       &  $\sim 50\%$ \\
\hline
\end{tabular}
\caption{Dominant branching fractions of various sparticles 
along the RNS model line for $m_{1/2}=1$~TeV.
\label{tab:BF}}
\end{center}
\end{table}
%

\section{Gluino cascade decay signatures}
\label{sec:gg}
%

We first examine the $pp\to\tg\tg X$ reaction followed by gluino cascade decays\cite{cascade} 
which can be searched for in multi-lepton plus multi-jet $+ \eslt$ events. 
We neglect squark pair production and gluino-squark associated
production which occur at very low rates because squarks are heavy.

We use Isajet~7.83~\cite{isajet} for the generation of signal events at LHC14.
For event generation, we use a toy detector simulation with calorimeter cell size
$\Delta\eta\times\Delta\phi=0.05\times 0.05$ and $-5<\eta<5$. The HCAL
(hadronic calorimetry) energy resolution is taken to be
$80\%/\sqrt{E}+3\%$ for $|\eta|<2.6$ and FCAL (forward calorimetry) is
$100\%/\sqrt{E}+5\%$ for $|\eta|>2.6$, where the two terms are combined
in quadrature. The ECAL (electromagnetic calorimetry) energy resolution
is assumed to be $3\%/\sqrt{E}+0.5\%$. We use the cone-type Isajet
 jet-finding algorithm~\cite{isajet}  to group the hadronic
final states into jets. Jets and isolated leptons are defined as follows:
\bi
\item Jets are hadronic clusters with $|\eta| < 3.0$,
$R\equiv\sqrt{\Delta\eta^2+\Delta\phi^2}\leq0.4$ and $E_T(jet)>50$~GeV.
\item Electrons and muons are considered isolated if they have $|\eta| <                          
2.5$, $p_T(l)>10 $~GeV with visible activity within a cone of $\Delta                             
R<0.2$ about the lepton direction, $\Sigma E_T^{cells}<5$~GeV.
\item  We identify hadronic clusters as
$b$-jets if they contain a B hadron with $E_T(B)>$ 15~GeV, $\eta(B)<$ 3 and
$\Delta R(B,jet)<$ 0.5. We assume a tagging efficiency of 60$\%$ and
light quark and gluon jets can be mis-tagged
as a $b$-jet with a probability 1/150 for $E_{T} \leq$ 100~GeV,
1/50 for $E_{T} \geq$ 250~GeV, with a linear interpolation
for intermediate $E_{T}$ values.
\ei

Gluino pair production cascade decay signatures have been previously
calculated and compared against backgrounds in Ref.~\cite{heaya}.  In that
paper, it was advocated that in models where gluino pair production
signatures are dominant above background (such as the focus point region
of mSUGRA), if one can suppress the background entirely, then the
remaining total cross section may be used to extract the gluino mass to
10-15\% precision.  We adopt the cuts from that paper and compare RNS signal rates
along the model line against previously calculated backgrounds using the
exact same set of cuts.

In Ref.~\cite{heaya}, the following pre-cuts set {\it C1} are first invoked\cite{frank}:
\\
\textbf{C1 Cuts:}
\bea
\eslt & >& \max (100\ {\rm GeV},0.2 M_{eff}),\nonumber \\
n(jets) &\ge & 4, \nonumber\\
E_T(j_1,j_2,j_3,j_4)& > & 100,\ 50,50,50\ {\rm GeV}, \label{c1cutsend}\\
S_T &>&0.2,\nonumber \\
p_T(\ell)& >& 20 \ {\rm GeV}.\nonumber
\eea
Here, $M_{eff}$ is defined as in Hinchliffe {\it et al.}\cite{frank} as
$M_{eff}=\eslt +E_T(j_1)+E_T(j_2)+E_T(j_3)+E_T(j_4)$, where $j_1-j_4$
refer to the four highest $E_T$ jets ordered from highest to lowest
$E_T$, $\eslt$ is missing transverse energy and $S_T$ is transverse
sphericity.  The SM cross sections in fb after C1 cuts are listed in
Table~III of Ref.~\cite{heaya}.  It is found that the signal with these
cuts is swamped by various SM backgrounds (BG), especially those from
QCD multi-jet production and $t\bar{t}$ production.  After inspection of
a variety of distributions including jet multiplicity $n(jets)$, $b$-jet
multiplicity $n(b-jets)$ and augmented effective mass $A_T$ (here,
$A_T=\eslt +\sum_{leptons}E_T +\sum_{jets}E_T$), for $0\ell$ and $1\ell$
events, we amend {\it C1} cuts to \\ \textbf{C2 Cuts:} \beas & apply\
cuts\ set\ C1 & \\ & n(jets)\ge 7 & \\ & n(b-jets)\ge 2 & \\ & A_T\ge
1400\ {\rm GeV}.  \eeas For multi-lepton events (opposite sign dileptons
$OS$, same sign dileptons $SS$ and trileptons $3\ell$), we use somewhat
softer cuts: \\ \textbf{C3 Cuts:} \beas & apply\ cuts\ set\ C1 & \\ &
n(isol.\ leptons)\ge 2 & \\ & n(jets)\ge 4 & \\ & n(b-jets)\ge 2 & \\ &
A_T\ge 1200\ {\rm GeV} .  \eeas

After {\it C2} cuts, it is found that 1~fb of BG remains in the $1\ell +jets$ channel and 
0.5~fb of BG remains in the $0\ell +jets$ channel. No BG was found in the 
$OS+jets$, $SS+jets$ or $3\ell +jets$ channels after cuts {\it C3}.
The signal rates along the RNS model line are shown in
Fig.~\ref{fig:multi}.
From the plot, we can read off the $5\sigma$ discovery level for
various integrated luminosity choices for different signal channels.
For the $0\ell +jets$ channel with $300$~fb$^{-1}$, we expect a reach to
$m_{1/2}\sim 650$~GeV corresponding to $m_{\tg}\sim 1.7$~TeV. We do not
project the reach in the lower background multilepton channels 
as these would depend on the residual background that remains. 

\FIGURE[tbh]{
\includegraphics[width=10cm,clip]{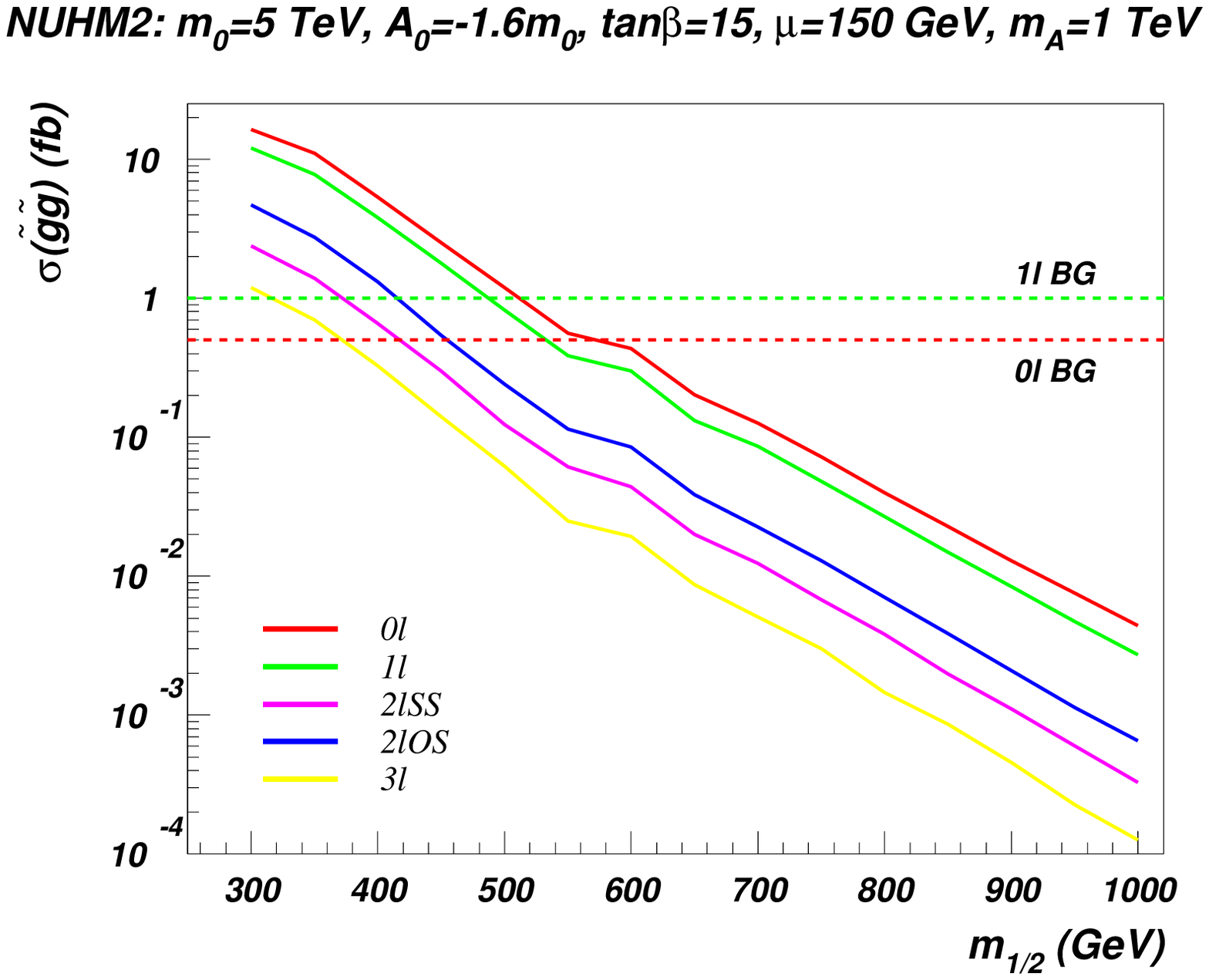}
\caption{Plot of gluino cross section in $fb$ after cuts {\it C2} for
$1\ell+jets$ and $0\ell+jets$ channel and cuts {\it C3} for $OS$, $SS$
and $3\ell+jets$ channels from gluino cascade decays along the RNS model
line at LHC14. The horizontal lines denote the corresponding backgrounds
estimated in Ref.~\cite{heaya}.  }
\label{fig:multi}}
%

\subsection{OS/SF dilepton mass distribution from cascade decays}
\label{ssec:mll}
%
Within the OS dileptons plus jets channel, we expect a large fraction of signal
events to contain an OS dilepton pair arising from
$\tz_2\to\ell^+\ell^-\tz_1$ decay. For these events, the $m(\ell^+\ell^-
)$ distributions will be bounded by the kinematic mass difference
$m_{\tz_2}-m_{\tz_1} < M_Z$.  In Fig.~\ref{fig:mll}, we show the invariant
mass of opposite-sign/same-flavor dilepton pairs from the $OS+jets$
events which survive cuts {\it C3}. In the figure we take
$m_{1/2}=450$~GeV for which $m_{\tg}=1250$~GeV and
$m_{\tz_2}-m_{\tz_1}=32$~GeV.  A mass edge at 32~GeV is clearly visible
from the plot, as is the $Z$ peak. A detection of an excess of events
with a cut-off on the dilepton mass could readily be attributed to
neutralinos of SUSY.
\FIGURE[tbh]{
\includegraphics[width=10cm,clip]{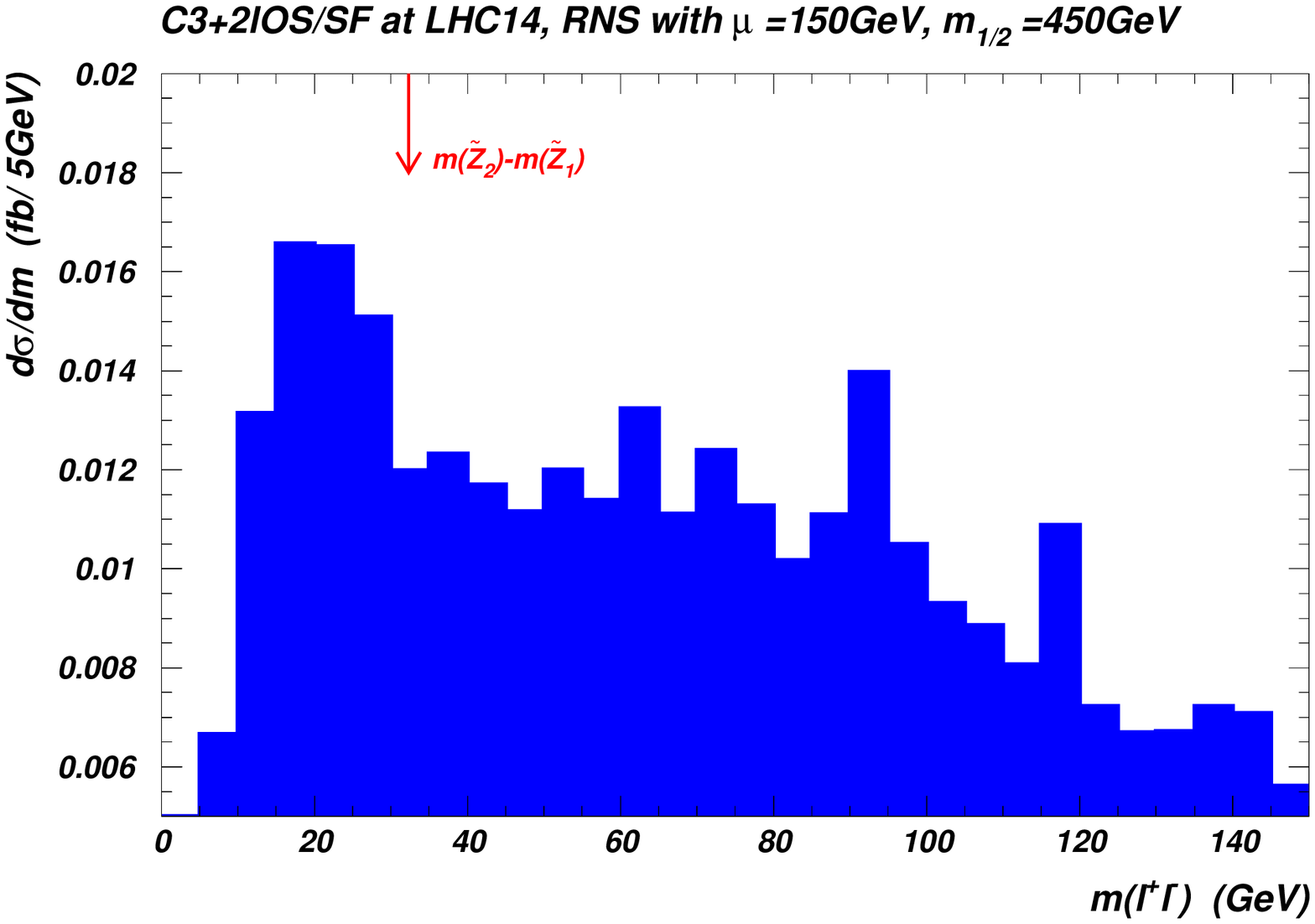}
\caption{Distribution of the invariant mass of opposite-sign/same-flavor
dileptons after cuts {\it C3} at LHC14 from the RNS benchmark model line
with $m_{1/2}=450$~GeV.}
\label{fig:mll}}
%

\section{Same-sign diboson signature}
\label{sec:SSdB}

In this Section, we revisit and present further details on the same-sign
diboson (SSdB) signature which was first introduced in
Ref.~\cite{lhcltr}. The idea here is that in models where $|\mu|$ is
smaller than the magnitude of gaugino mass parameters -- as exemplified
by the RNS model line -- wino pair production provides a
novel signature with a final state characterized by two same sign $W$
bosons and $\eslt$ but accompanied by just modest jet activity. 
The most promising reaction appears to be $pp\to \tw_2^\pm\tz_4$, where $\tw_2^\pm\to
W^\pm\tz_{1,2}$ and $\tz_4\to W^\pm\tw_1^\mp$ although $\tw_2^+\tw_2^-$ pair
production also provides a non-negligible signal contribution. 
We see from Fig.~\ref{fig:bf} that the winos have substantial branching 
fractions for decays to $W$ bosons.
For these decays, half the time the final states consist of $W^\pm W^\pm +\eslt$. 
We focus our attention on the SS dilepton signal from the leptonic decays of both $W$s.
The jet activity in these events is relatively limited  since 
the daughter higgsinos $\tw_1$ and $\tz_{1,2}$ usually yield only soft decay products. 
This serves to distinguish the wino-pair induced SSdB signature from the
SS dilepton signal from gluino pair production -- the latter is expected to be accompanied by
several hard jets. 

The SM physics backgrounds to the SSdB signal come from $uu\to            
W^+W^+ dd$ or $dd\to W^-W^- uu$ production with a cross section $\sim 350$~fb. 
These events will be characterized by high rapidity
(forward) jets and rather low $\eslt$. $W^\pm W^\pm$ pairs may also occur
via two overlapping events: such events will mainly have low $p_T$ $W$s
and possibly distinct production
vertices. Double parton scattering will also lead to SSdB events at a
rate somewhat lower than the $qq \to W^\pm W^\pm q'q'$ process\cite{stirling}. 
Additional physics backgrounds come from
$t\bar{t}$ production where a lepton from a daughter $b$ is
accidentally not isolated, from $t\bar{t}W$ production and from $4t$ production. 
SM processes such as $WZ\to 3\ell$ and $t\bar{t}Z\to 3\ell$ production,
where one lepton is missed, constitute {\em reducible} backgrounds to the
signal.

Here, we assume that the $2\to 4$ processes as well as the double parton
scattering processes, which have different characteristics from the
signal, can be readily eliminated by suitable cuts and do not simulate
these.  For the simulation of the remaining background events we use
AlpGen~\cite{alpgen} and MadGraph~5~\cite{madgraph} to generate the hard
scattering events. Those events are then passed to Pythia~6.4~\cite{pythia}
via the LHE interface~\cite{lhe} for showering and hadronization.  For 
the $2\to 4$ ``$WZ$''
process, we compute the full matrix element for $pp\rightarrow l^+ l^-
l' \nu'$ that includes contributions from on- and off-shell $Z$
and $\gamma$ as well as from interference diagrams.  We normalize
signal and background to NLO cross sections obtained with
Prospino~\cite{prospino} and MCFM~\cite{mcfm}, respectively. To
reconstruct jets and isolated leptons we followed the procedure described in
Sec.~\ref{sec:gg}.

In Ref.~\cite{lhcltr}, the following cuts were imposed:
\bi
\item {\it exactly} 2 isolated same-sign leptons with 
$p_T(\ell_1 )>20$~GeV and $p_T(\ell_2 )>10$~GeV,
\item $n(b-jets)=0$ (to aid in vetoing $t\bar{t}$ background).
\ei
After these cuts, the event rate is dominated by $WZ$ and $t\bar{t}$
backgrounds.  

To distinguish signal from background, we next construct the transverse
mass of each lepton with $\eslt$: 
\be 
\mtmin\equiv \min\left[
m_T(\ell_1,\eslt),m_T(\ell_2 ,\eslt )\right] \nonumber \, .  
\ee 
The signal gives rise to a continuum distribution, whilst the dominant
backgrounds have a kinematic cut-off around $\mtmin \simeq M_W$ (as long
as the $\eslt$ dominantly arises from the leptonic decay of a single
$W$).  The situation is shown in Fig.~\ref{fig:ssdist}, where we show in
{\it a}) the $\mtmin$ distribution, while in {\it b}) we show the
$\eslt$ distribution.  The bulk of $t\bar{t}$ and $WZ$ backgrounds can
be eliminated by requiring
\bi
\item $\mtmin >125$~GeV and
\item $\eslt >200$~GeV.  
\ei 
After these cuts, we are unable to generate any background events from
$t\bar{t}$ and $WZ$ production, where the 1-event level in our
simulation was 0.05~fb and 0.023~fb, respectively.  The dominant SM
background for large $\mtmin$ then comes from $Wt\bar{t}$ production for
which we find (including a QCD $k$-factor $k=1.18$ obtained from
Ref.~\cite{garzelli}) a cross section of $0.019$ ($0.006$)~fb after the
cuts $\mtmin > 125$~(175)~GeV and $\eslt>200$~GeV; the harder cuts serve
to optimize the signal reach for high $m_{1/2}$ values.\footnote{We have
ignored detector-dependent backgrounds from jet-lepton misidentification
in our analysis, but are optimistic that these can be controlled by the
$\mtmin$ and $\eslt$ cuts. Estimates of the background from charge
mis-identification, which could be important, especially for electrons, are
also beyond the scope of this analysis.}
\FIGURE[tbh]{
\includegraphics[width=8cm,clip]{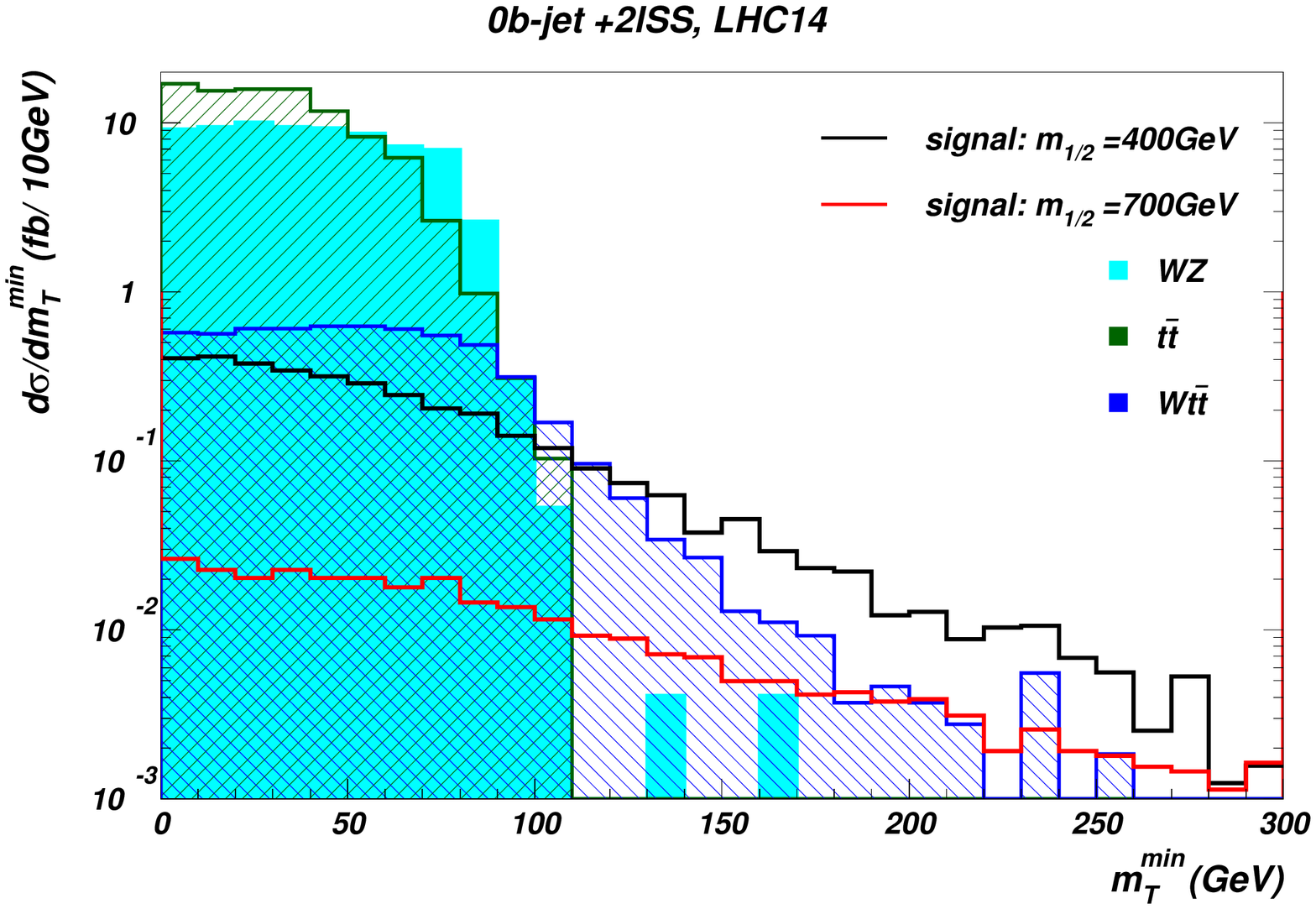}
\includegraphics[width=8cm,clip]{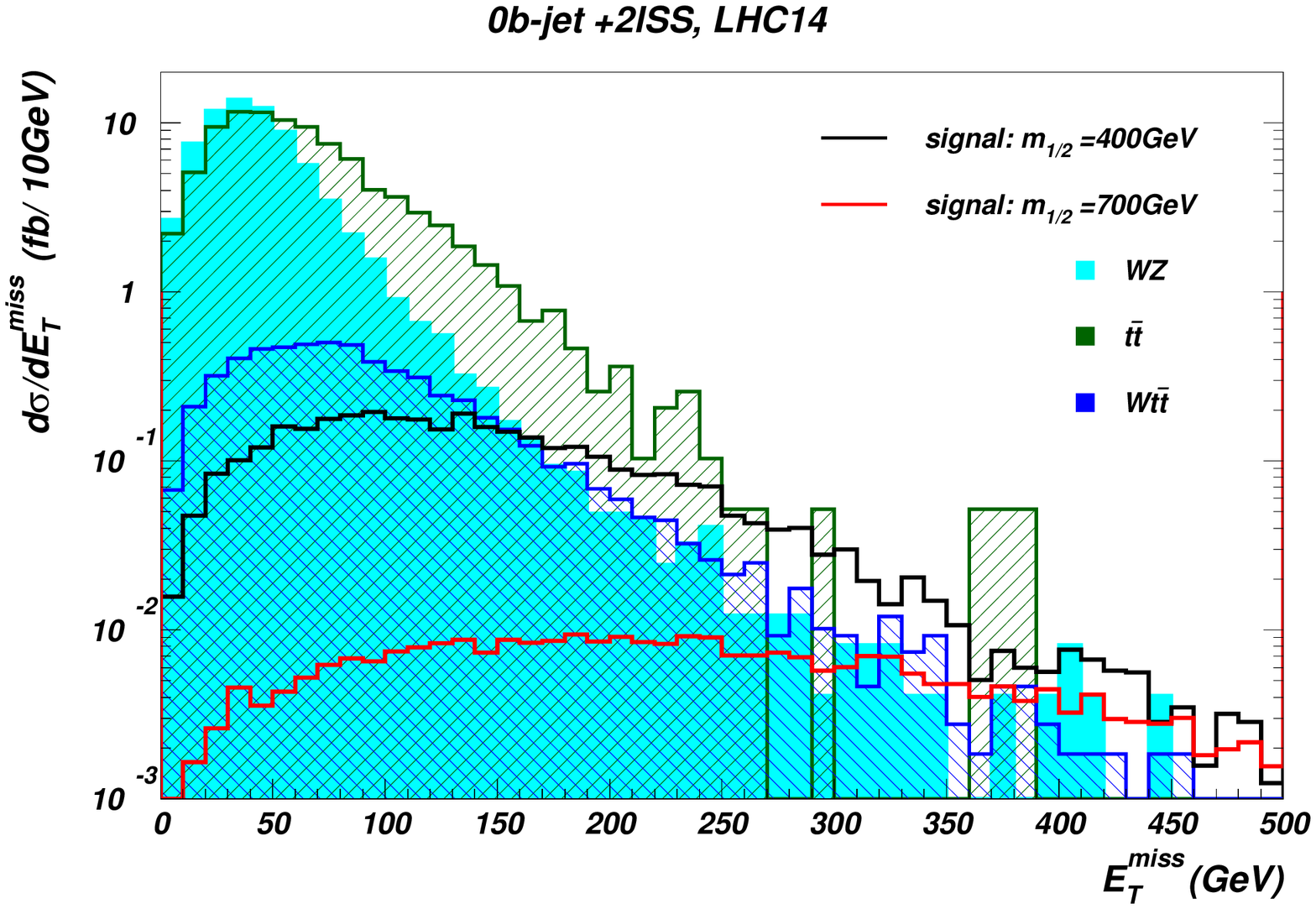}
\caption{Transverse mass and missing energy distributions for SSdB 
events after cuts at LHC14.  The open black and red histograms
represent the signal from winos -- via $\tw_2\tz_4$ and $\tw_2^+\tw_2^-$
pair production -- for the RNS model-line points with $m_{1/2}=400$~GeV and 700~GeV, respectively.}
\label{fig:ssdist}}

The calculated signal rates after cuts along the RNS model line from
just $\tw_2^\pm\tz_4$ and $\tw_2^\pm\tw_2^\mp$ production are shown
vs. $m_{1/2}$ in Fig.~\ref{fig:ss} where the upper (blue) curves require
$\mtmin>125$~GeV and the lower (orange) curve requires $\mtmin>175$~GeV.
The $\tw_2\tz_4$ and $\tw_2\tw_2$ cross sections are normalized to those
from Prospino\cite{prospino}.  For observability with an assumed value
of integrated luminosity, we require: 1)~significance $> 5\sigma$,
2)~Signal/BG$>0.2$ and 3)~at least 5 signal events.  The LHC signal
(blue dashed curve) and reach lines for integrated luminosity values 25
and 100~fb$^{-1}$ with a soft $\eslt >75$~GeV cut are shown first. The
25~fb$^{-1}$ reach is to $m_{1/2}\simeq 450$~GeV corresponding to
gluinos of $\sim 1300$~GeV. As greater integrated luminosity is
accumulated, harder cuts can be applied. The solid blue line shows signal
for $\eslt >200$~GeV and reach for 100, 300 and 1000~fb$^{-1}$.  With
harder cuts, the 100~fb$^{-1}$ reach extends to $m_{1/2}\simeq 680$~GeV
corresponding to $m_{\tg}\sim 1.75$~TeV in a model with gaugino mass unification. 
The direct search for $\tg\tg$ gives a projected reach of
$m_{\tg}\sim 1.6$~TeV as we have seen in Sec.~\ref{sec:gg}; see also
Ref.~\cite{bblt2012}. Thus, with ${\cal O}(100)$~fb$^{-1}$ of integrated
luminosity, the SS diboson
signal offers a comparable reach to that for gluino cascade decays.  For
300~(1000)~fb$^{-1}$ of integrated luminosity, the reach is improved
with a harder $\mtmin >175$~GeV cut. In this case, we find the LHC14
reach for SS dibosons extends to $m_{1/2}\sim 840$~(1000)~GeV,
corresponding to $m_{\tg}$ of 2.1 and 2.4~TeV.  For the RNS model-line
where gaugino mass unification is assumed, these reach numbers extend
well beyond the LHC14 reach for direct gluino pair
production\cite{bblt1014}. Regardless of this, we emphasize that the
SSdB signal is a new independent signal, and detection of
signals in multiple channels will be essential to unravel the underlying
origin of any new physics that is found. 
\FIGURE[tbh]{
\includegraphics[width=10cm,clip]{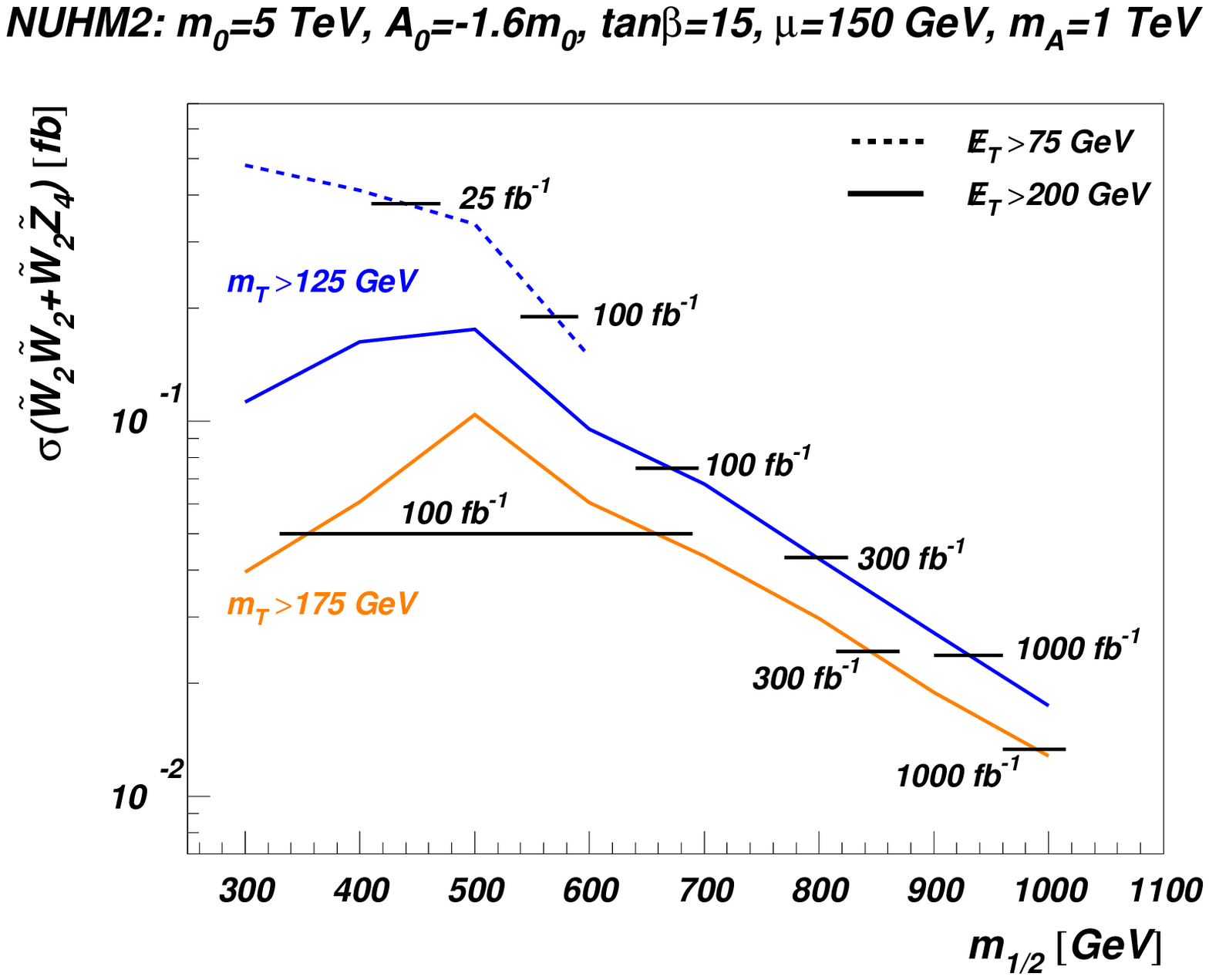}
\caption{Same-sign dilepton cross sections (in {\it fb}) at LHC14 after cuts vs. $m_{1/2}$
along the RNS model line from $\tw_2^\pm\tz_4$ and
$\tw_2^\pm\tw_2^\mp$ production and
calculated reach for 100, 300 and 1000~fb$^{-1}$.  The upper solid and
dashed (blue) curves requires $\mtmin >125$~GeV while the lower solid
(orange) curve requires $\mtmin >175$~GeV. The signal is observable
above the horizontal lines.
}
\label{fig:ss}}

We stress again that the low jet activity associated with the SSdB
signal from SUSY models with light higgsinos makes it quite distinct
from the usual SS dilepton signal arising from gluino pair production,
which is usually accompanied by numerous hard jets and high $\eslt$.
Recent CMS searches for SS dileptons from SUSY\cite{cms_ss} required the
presence of multiple jets (some $b$-tagged jets) or large $H_T$ in the
events; these cuts greatly reduce or even eliminate our SSdB
signal. Likewise, the cuts $n_j\ge 3$ high $p_T$ jets (possibly
$b$-tagged) along with $\eslt>150$~GeV and large $m_{\rm eff}$ required
by a recent ATLAS search for SS dileptons from gluinos\cite{atlas_ss}
would have eliminated much of the SSdB signal from SUSY with light
higgsinos.

Hard trilepton production from winos (discussed in the next section) can
lead to clean, same-sign dilepton events if a lepton is not isolated or
fails to be identified. The CMS collaboration used this channel to
extend the search for electroweak-inos to portions of parameter space
not accessible via the trilepton search, requiring 120~GeV $< \eslt <
200$~GeV \cite{cmsdilep}.  They do not, however, impose the $m_T^{\rm
min}$ cut that we found crucial for our SSdB analysis. The CMS search is
thus not optimized for the clean SS dilepton signal in the RNS
scenario. In any case, with just $\sim 20$~fb$^{-1}$ at LHC8, this
channel should have a lower reach than that via multi-jet plus
multi-lepton events from gluino pair production.

\section{Hard trileptons from wino pair production}
\label{sec:WZ}

In this Section, we examine prospects for detection of reactions such as
\be 
pp\to\tw_2\tz_4\to (\tw_1 Z)+(\tw_1 W)\to WZ+\eslt\to
\ell^+\ell^-\ell^\prime+\eslt. \nonumber\ 
\ee 
The trilepton channel
where the neutralino decays via the three-body decay $\tz_2\to
\ell^+\ell^-\tz_1$ because the two-body decay $\tz_2\to Z\tz_1$ is
kinematically forbidden (so that SM trileptons from $WZ$ production can
be eliminated via a mass cut on the opposite-sign, same flavour dilepton
pair) has long been regarded as a golden channel in the search for
gauginos from supersymmetry~\cite{trilep}.
More recently, it has been pointed out~\cite{wz} that at least within
mSUGRA the trilepton search for gauginos is viable even when the
neutralinos decay to on-shell $Z$ bosons. Indeed, the CMS and ATLAS
experiments have searched in this channel and found that there is no
excess above SM expectations~\cite{lhctrilepZ}.  For a recent assessment
of multilepton signals, see Ref.~\cite{han}. 
Here, we analyse prospects for this signal for the RNS model line 
for the most part 
following the cuts of Ref.~\cite{wz} which required:
\bi
\item[] \underline{Pre-Selection Cuts:}
\item $n(b-jets)=0$ (to aid in vetoing $t\bar{t}$ background),
\item 3 isolated leptons with $p_T(\ell )>20$~GeV and
\item $|m(\ell^{+}\ell^{-}) - M_Z| < 10$~GeV (leptonic $Z$), 
\ei 
where two of the leptons in the event must form an OS/SF pair. If more
than one OS/SF pairing is possible, the pair which minimizes
$|m(\ell^{+}\ell^{-}) - M_Z|$ is chosen.  The remaining lepton is
labeled $\ell^{'}$.  In the case of the RNS model line, the $WZ+\eslt$ signal
also receives a smaller, though non-negligible contribution, from
$\tw_2^+\tw_2^-$ where one of the winos decays via $\tw_2\to W\tz_{1,2}$ and
the other via $\tw_2\to Z\tw_1$ mode. 
At this point, a large background from the $2\to 4$
process $pp\to (\ell^+\ell^-)+(\ell^{\pm\prime}\nu_{\ell^\prime})$ which
occurs via various on- and off-shell processes -- including $W^*Z^*$ and
$W^*\gamma^*$ production -- tends to dominate the signal. Here, we
re-evaluate the $2\to 4$ process using MadGraph with no restriction on
the invariant mass around the $Z$ and $W$ resonances.\footnote{The $WZ$
background after all cuts in Table~1 of Ref.~\cite{wz} has been
underestimated by a factor of about 2.5 because the virtual $W$-mass
did not extend to large enough values, leading to an under-estimate of
the tail of the $m_T$ distribution.} For $t\bar{t}$, $Z(ll)+jets$,
$W(l\nu)+jets$, $Z(ll)+t\bar{t}$ and $W(l\nu)+t\bar{t}$ (all summed over 3
lepton flavors) we allow for at least two additional partons in the
final state and use the MLM matching scheme~\cite{mlm} to avoid double
counting.  We have also included $ZZ$, $W(l\nu)+tb$ and $Z(ll)+b\bar{b}$
backgrounds.  The signal and background distributions in
$m_{T}(\ell^{'},\eslt )$ and $\eslt$ are shown in Fig.~\ref{fig:mT}.
\FIGURE[tbh]{
\includegraphics[width=8cm,clip]{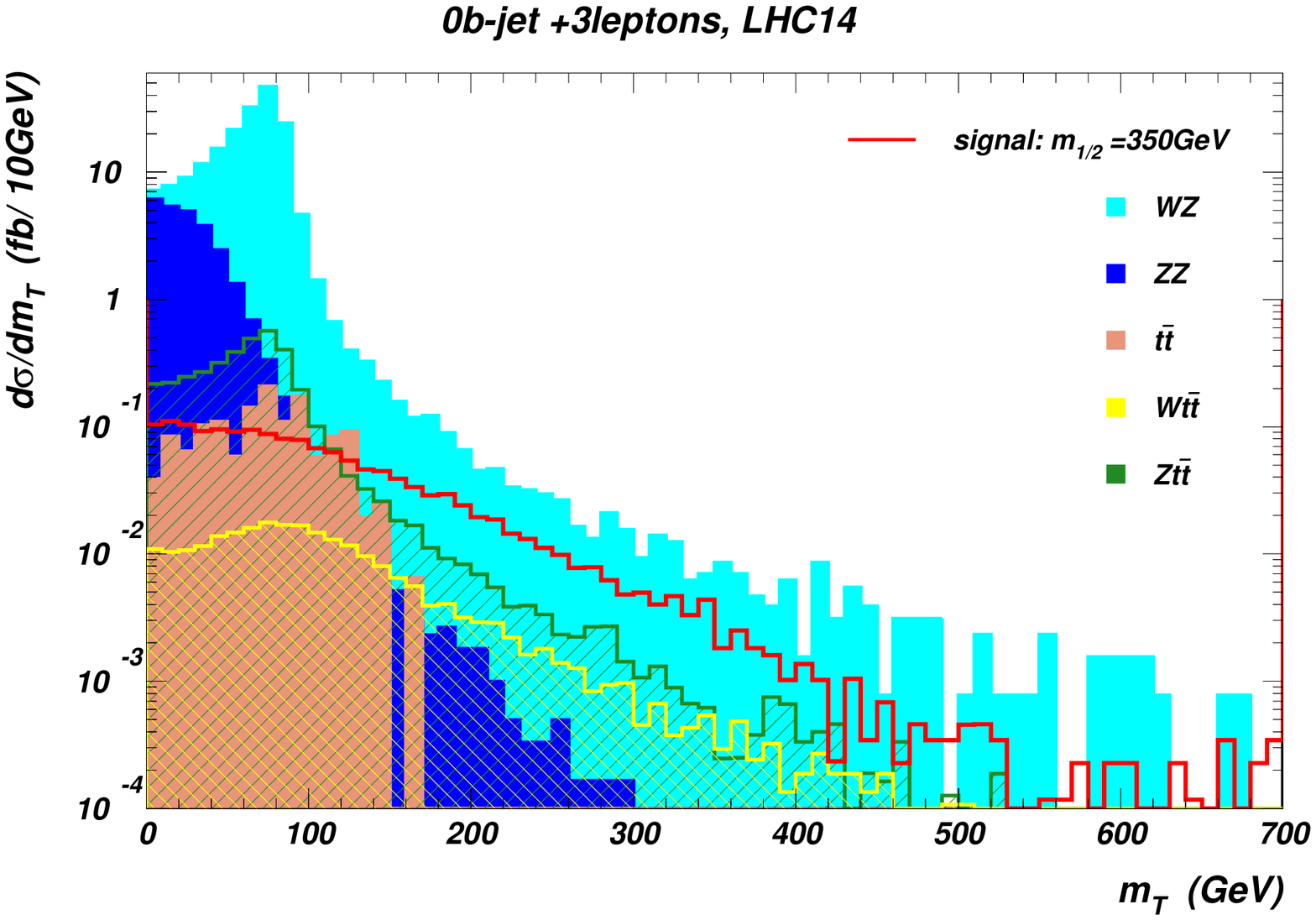}
\includegraphics[width=8cm,clip]{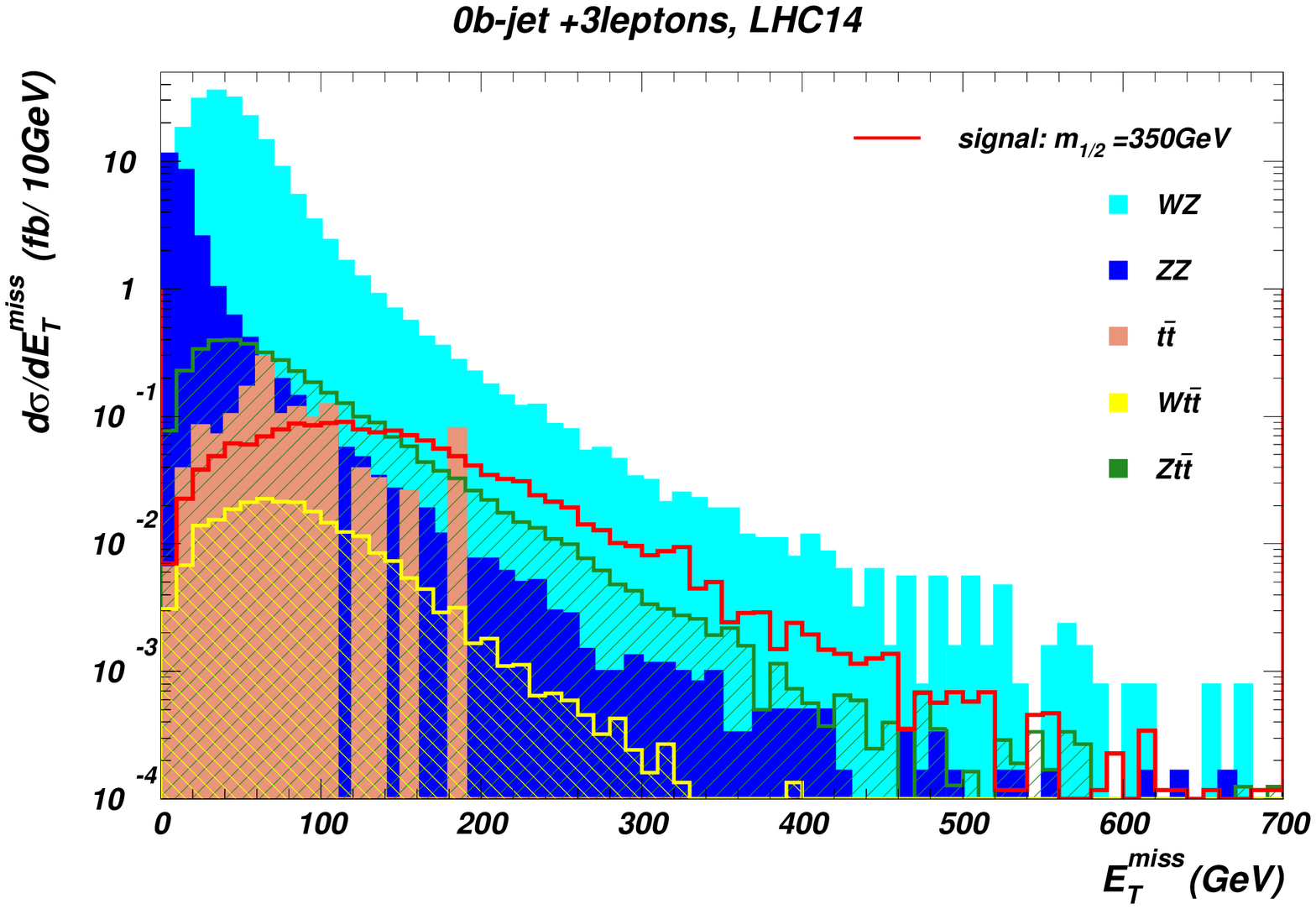}
\caption{Transverse mass and missing energy distributions for hard trilepton
  events after the preliminary cuts at LHC14.  The open red histograms
  represent the signal from winos, $\tw_2$ and $\tz_4$, for the RNS
  point with $m_{1/2}=350$~GeV.  }
\label{fig:mT}}

To enhance the signal relative to background, we then require,
\bi
\item $m_{T}(\ell^{'},\eslt ) > 125$~GeV,
\item $\eslt >150$~GeV.
\ei
The $m_T$ cut is as in Ref.~\cite{wz}, but for the larger integrated
luminosity and concomitantly higher wino masses that we are considering
here, we found that stiffening the $\eslt$ cut yields a better
signal-to-background ratio.  The background from various SM sources
along with RNS signal for $m_{1/2}=350$~GeV is shown after cuts in
Table~\ref{tab:WZ} for LHC14.
\begin{table}
\begin{center}
\begin{tabular}{|l|r|r|r|r|r|r|r|}
\hline
& $t\bar{t}$ & $WZ$ & $ZZ$ & $Z+t\bar{t}$ &  $W+t\bar{t}$ & Total BG & Signal\\
\hline
Events Generated & 12M & 1.5M & 1M & 1.2M & 10M & & 200K \\
\hline
$n(b) = 0, n(l) = 3$  & 6.96 & 211.94 & 26.07 & 4.26 & 1.84 & 247.29 & 2.88 \\
OS/SF pair            & 5.25 & 211.51 & 26.02 & 4.21 & 1.37 & 251.97 & 2.57 \\
$m(\ell^{+}\ell^{-})$ cut & 0.95 & 186.90 & 25.55 & 3.99 & 0.24 & 221.20 & 1.52 \\
$m_{T} > 125$~GeV   & 0.03 & 1.64 & 0.05  & 0.20 & 0.07 & 1.99 & 0.43 \\
$\eslt > 150$~GeV    & 0.006 & 0.24 & $< 0.00085$ & 0.0058 & 0.016 & 0.32 & 0.22 \\
\hline
\end{tabular}
\caption{Number of events generated and
cross section after cuts for the dominant backgrounds in the hard trilepton
channel and for the RNS signal with $m_{1/2}=350$~GeV. 
All cross sections are in {\it fb}. 
The total BG values include all processes listed in the text, including the
subdominant ones not shown in the Table.
\label{tab:WZ}}
\end{center}
\end{table}

In Fig.~\ref{fig:WZ}, we show the $3\ell +\eslt$ signal cross section after
all cuts versus $m_{1/2}$ along the RNS model line. The turn-over at
the left end of the curve is because of the efficiency loss resulting 
from the stiff $\eslt$ cut which is optimized to yield the best
reach for high wino masses. For 100~fb$^{-1}$ of
integrated luminosity, there is no reach, while the reach in $m_{1/2}$
is shown for 300 and 1000~fb$^{-1}$. The 300~fb$^{-1}$ reach extends to
$m_{1/2}=500$~GeV while the 1000~fb$^{-1}$ reach extends to
$m_{1/2}=630$~GeV. 
These values correspond to gluino masses of
$m_{\tg}=1.3$~TeV and 1.65~TeV, respectively. These reaches are smaller
than those obtained from the $\tg\tg$ and SSdB signals. They
would, nevertheless, offer corroborative evidence for any SUSY discovery at the
lower range of allowed $m_{1/2}$ values.
\FIGURE[tbh]{
\includegraphics[width=10cm,clip]{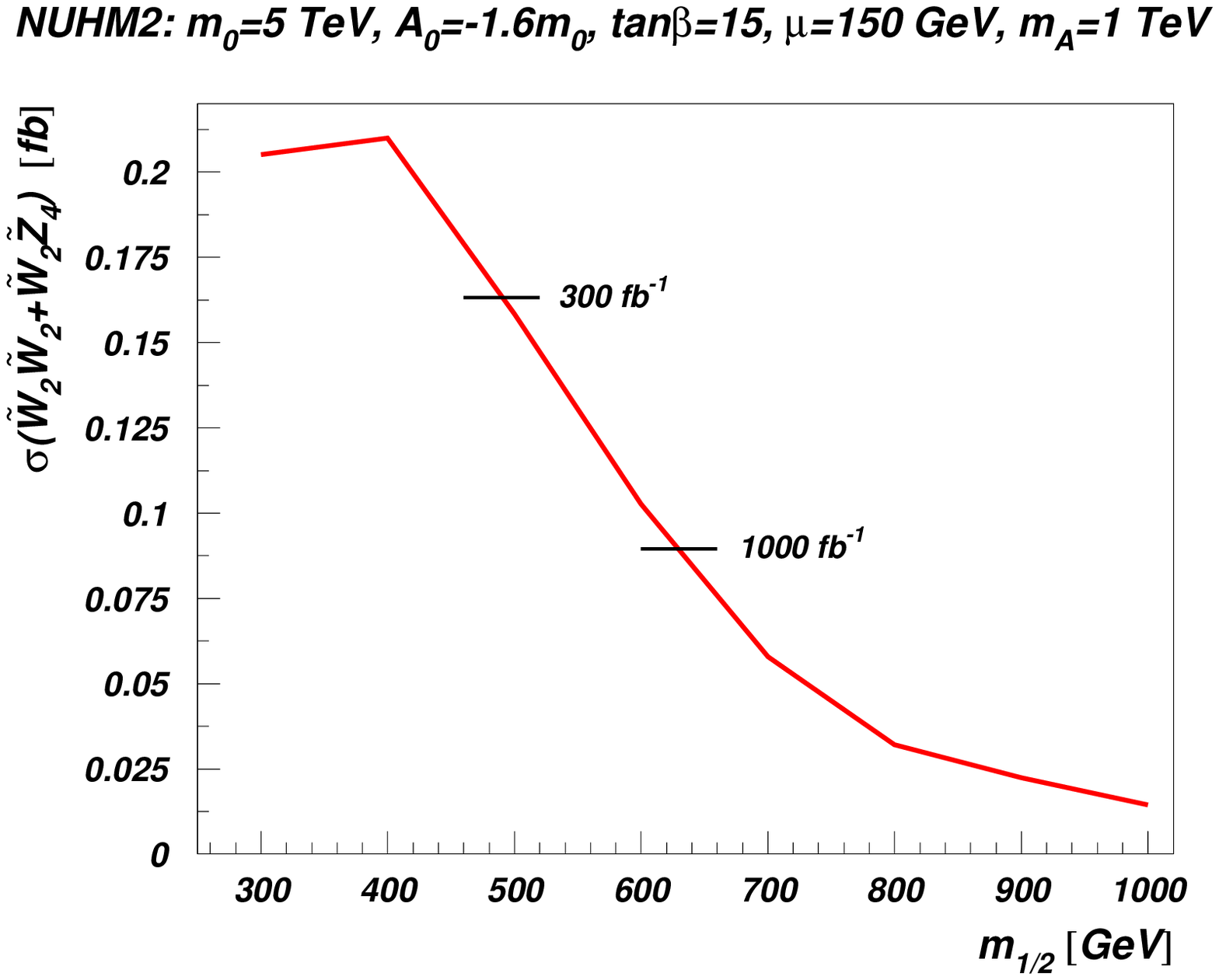}
\caption{Tri-lepton cross sections (in {\it fb}) at LHC14 after cuts
vs. $m_{1/2}$ along the RNS model line from wino pair production processes
$pp\to\tw_2\tz_4 / \tw_2\tw_2 \to WZ+\eslt\to 3\ell+\eslt$ events.  }
\label{fig:WZ}}

\section{Four leptons from heavy gaugino production} 
\label{sec:4l}

We saw in Fig.~\ref{fig:bf} that the wino-like $\tw_2$ and $\tz_4$ have
significant branching fractions to $W$ and $Z$ bosons resulting in the
dilepton and trilepton signals already discussed. A small fraction of
the time, there may be two $Z$ bosons in these events, leading to the
possibility of a four-lepton signal. Additional leptons can arise from
the leptonic decays of daughter $\tw_1$ and $\tz_2$. Although the decay
products are generally soft, the ubiquity of these light higgsino-like
states within the RNS framework often results in additional detectable
leptons ($e$ and $\mu$) in would-be trilepton events. This
characteristic feature of low $|\mu|$ models such as RNS is absent in
models such as mSUGRA that have received the most attention in the
literature, and leads to the possibility of four-lepton plus $\eslt$
signal, even in $R$-parity conserving SUSY.\footnote{It is well-known
that high lepton multiplicities are obtained if the LSP decays via
lepton-number-violating interactions that do not conserve $R$-parity.}
A study of this new signal for which we require
\bi
\item 4 isolated leptons with $p_T(\ell)> 10$~GeV within $|\eta(\ell)|<
  2.5$,
\item $n_b=0$, to veto backgrounds from top decays,
\item $\eslt > \eslt({\rm cut})$, where $\eslt({\rm cut})$ is chosen
  to select signal events above SM backgrounds,
\ei
forms the subject of this section.
\FIGURE[tbh]{
\includegraphics[width=10cm,clip]{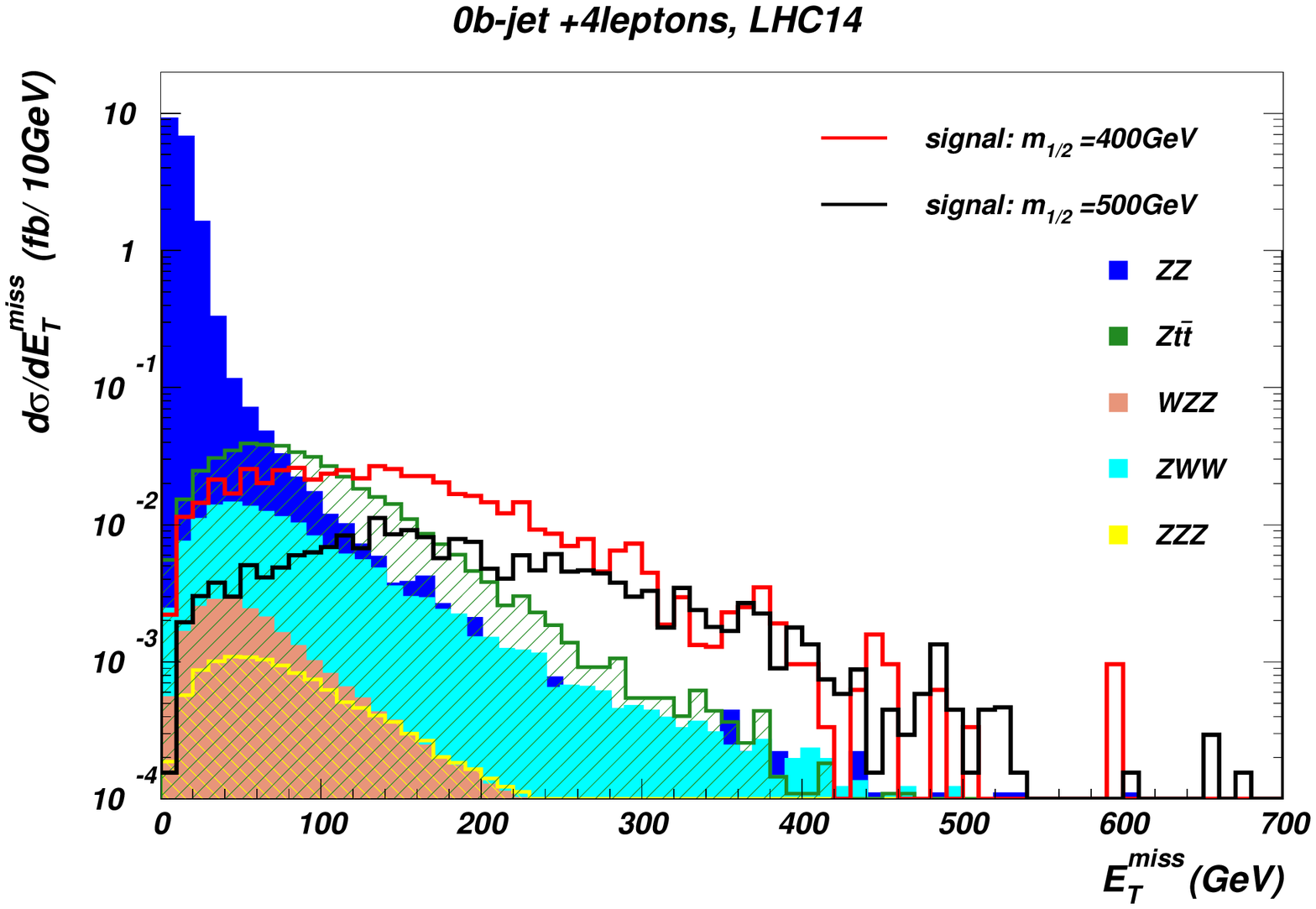}
\caption{The $\eslt$ distributions for $4\ell$ events with $n_b=0$ from 
various SM sources and for two signal points on the RNS model-line.  }
\label{fig:4l}}

Within the SM, the main sources of $4\ell + \eslt$ events are
$ZZ$, $Zt\bar{t}$, $ZWW$, $ZZW$, $ZZZ$ and $Zh(\to WW^*)$,
followed by leptonic decays of tops, and of the electroweak vector bosons. 
The bulk of
the background from $ZZ$ production is eliminated by requiring a large
$\eslt$. Nevertheless, this background remains significant since $\eslt$
can arise via $Z\to \tau^+\tau^- \to \ell^+\ell'^- +\eslt$.

We have simulated the RNS signal along with backgrounds from $ZZ$,
$t\bar{t}Z$ and $VVV$ ($V=W,Z$) using AlpGen and Pythia. 
The cross sections for the most important of these backgrounds
are listed in the second column of Table~\ref{tab:4l}, together with that for the
signal for three model-line points. The last two columns list these
signal and background calculations for $\eslt({\rm cut}) = 100$ and
200~GeV, the choice being motivated by the $\eslt$ distributions shown
in Fig.~\ref{fig:4l}. The numbers in bold-face show the statistical
significance for an integrated luminosity of 300~fb$^{-1}$.
\begin{table}
\begin{center}
\begin{tabular}{|l|c|c|c|}
\hline
cuts & $n(b) = 0, n(l) = 4$ & $\eslt > 100$~GeV & $\eslt > 200$~GeV \\
\hline
\hline
$ZZ$        & 18.02 & 0.0611 & 0.0094 \\
$Zt\bar{t}$ & 0.450 & 0.158  & 0.0232 \\
$ZWW$       & 0.155 & 0.0516 & 0.0134 \\
\hline
Total BG    & 18.66 & 0.280  & 0.0483 \\
\hline
$m_{1/2}=400$~GeV & 0.527 & 0.343 \ ({\bf 11.2}) & 0.122 \ ({\bf 3.8})\\
$m_{1/2}=500$~GeV & 0.195 & 0.157 \ ({\bf 5.1}) & 0.0769 \ ({\bf 6.1}) \\
$m_{1/2}=600$~GeV & 0.084 & 0.0728 \ ({\bf 2.3}) & 0.0467 \ ({\bf 3.7}) \\
\hline
\end{tabular}
\caption{Background and signal rates in $fb$ for 4-lepton events at
LHC14 after cuts. The bold-faced numbers in parenthesis in the last two
columns show the statistical significance of the signal with
300~fb$^{-1}$ of integrated luminosity at LHC14. The signal comes from
wino pair production for points on the RNS model line introduced in the text.
\label{tab:4l}}
\end{center}
\end{table}

We estimate that $Zh(\to W^\pm \ell\nu)$ yields a $4\ell$ cross section
 $\sim 1300$~fb $\times 0.06 \times 0.22 \times 0.03$ (where the last
 factor is the branching fraction for $h\to W^\pm \ell^\mp\nu$ decay)
 $\simeq 0.5$~fb, before any lepton acceptance cuts which further reduce
 the cross section by factor about 5-10. After the hard $\eslt$ requirement, we
 expect this to make a relatively unimportant contribution to the
 background. Backgrounds from $ttWW$ and $4V$ processes should also be
 small.

Several comments are worth noting.
\bi
\item There is no benefit, and in fact a loss of significance, by requiring pairs of
  leptons to reconstruct to $M_Z$. This is largely because the largest
  $4\ell$ backgrounds also all have a $Z$ in them, and both signal and
  backgrounds drop roughly equally due to this requirement. 

\item The softer $\eslt > 100$~GeV cut works better for
  $m_{1/2}=400$~GeV for which a 6$\sigma$ signal is obtained
  even with just  100~fb$^{-1}$ of integrated luminosity.

\item The $5\sigma$ reach for 300~fb$^{-1}$ (1000~fb$^{-1}$) extends
 to  $m_{1/2} = 500$~GeV (beyond $m_{1/2}=600$~GeV) with the harder
  $\eslt({\rm cut})=200$~GeV.
\ei 
We conclude that the $4\ell$ channel
  would serve to confirm a the SSdB signal pointing to light higgsinos
  out to $m_{1/2}$ values $\lesssim500-650$~GeV, depending on the
  integrated luminosity that is ultimately available.

We remark that current ATLAS~\cite{atlas_4l} and CMS~\cite{cms_4l} 4-lepton searches 
are optimized for the signal from the cascade decays of gluinos 
(and so do not veto hadronic activity)  
with the high lepton multiplicity originating in R-parity violating leptonic decays of $\tz_1$. 
In contrast, our signal is hadronically quiet and would stand out over SM backgrounds with veto
on b-jets as described in the text.

\section{Soft trileptons from direct higgsino pair production}
\label{sec:3l}

In this Section, we try to exploit the large cross sections for higgsino
pair production from the RNS model at the LHC:
$pp\to\tw_1\tz_1,\ \tw_1^+\tw_1^-,\ \tz_1\tz_2$ and $\tw_1^\pm\tz_2$.
The purely hadronic$+\eslt$ final states from higgsino pair production
are expected to be buried beneath prodigious QCD backgrounds since the
signal yields only soft, low $p_T$ jets and soft $\eslt$
spectra. Likewise, most single and dilepton signals are expected to be
buried under $W\to \ell\nu_\ell$ and $WW$, $t\bar{t}$ backgrounds
respectively.

In previous related work, Ref.~\cite{bkpu} 
did find a reach for mixed higgsino-gaugino $\tw_1\tz_2\to 3\ell$ 
(but not better than that via the gluino search) 
along the focus point region of
mSUGRA at LHC14, where the dominant background came from $t\bar{t}$ production, as opposed
to studies for the Tevatron, where $W^*Z^*\to (\ell\nu_\ell)+(\ell'\bar{\ell}')$ 
was dominant\cite{tev3l}. In the FP study of Ref.~\cite{bkpu}, the mixed gaugino-higgsino
like $\tw_1$ and $\tz_2$ had larger mass gaps, and hence harder decay products for the signal,
as compared with the present case.
Meanwhile, in Ref.~\cite{bbh}, the light higgsino-world scenario was investigated, with a 
focus on the reaction $pp\to\tz_1\tz_2\to\ell^+\ell^- +\eslt$. In that study, large backgrounds from
$WW$, $\gamma^*\to\ell^+\ell^-$ and $\tau\bar{\tau}$ production thwarted the signal 
of low-mass, collimated OS/SF dileptons, even with $p_T$ cuts extending down to 5~GeV. 
Here, we examine the clean trilepton signal from higgsino pair production.
We generate signal events for several points along the RNS model line. 

We will search for the $pp\to\tw_1\tz_2\to (e\nu_e\tz_1)+(\mu^+\mu^-\tz_1)$ topology where we
assume a dilepton trigger with $p_T(e)>10$~GeV and $p_T(\mu )>5$~GeV. Then we require:
\bi
\item $10\ {\rm GeV}<p_T(e)<50$~GeV, 
\item $5\ {\rm GeV}<p_T(\mu_{1})<50$~GeV,
\item $5\ {\rm GeV}<p_T(\mu_{2})<25$~GeV.
\ei
Scrutiny of a variety of distributions suggests the following cuts:
\begin{enumerate}
\item $10\ {\rm GeV}<m(\mu^+\mu^- )<75$~GeV,
\item $n(jets) =0$ (jet-veto),
\item electron transverse mass $m_T(e ,\eslt )<60$~GeV,
\item $25\ {\rm GeV}<\eslt <100$~GeV.
\end{enumerate}

The signal and four background processes are shown in Table~\ref{tab:3l}. 
\begin{table}
\begin{center}
\begin{tabular}{|l|r|r|r|r|r|}
\hline
cuts & $t\bar{t}$ & $W^*Z^*$ &  $ZZ$ & $Wt\bar{t}$ & signal \\
\hline
\hline
cut 1  & 12.4 & 7.6  & 0.15 & 0.1 & 0.42 \\
cut 2  & 2.4  & 7.1  & 0.09 & 0.006 & 0.42 \\
cut 3  & 1.3  & 4.4  & 0.08 & 0.003 & 0.42 \\
cut 4  & 0.9  & 2.0  & 0.03 & 0.002 & 0.28 \\
\hline
\end{tabular}
\caption{Background and signal rates in $fb$ for soft $3\ell +\eslt$ 
events at LHC14 after cuts. The signal comes from higgsino pair production at $m_{1/2}=400$~GeV point 
on the RNS model line. The $2\to 4$ process labelled $W^*Z^*$ includes $\gamma^*,\ Z^*\to\tau\bar{\tau}$.
\label{tab:3l}}
\end{center}
\end{table}
In Fig.~\ref{fig:softe} we show distributions of electron $p_T$ before cuts for two sample 
points on the RNS model line. We see that energy release is very small, less than $\sim
25$~GeV, and quickly decreases with $m_{1/2}$. For $m_{1/2}=1$~TeV most of electrons have
$p_T$ less than 10~GeV, the trigger threshold.
\FIGURE[tbh]{
\includegraphics[width=10cm,clip]{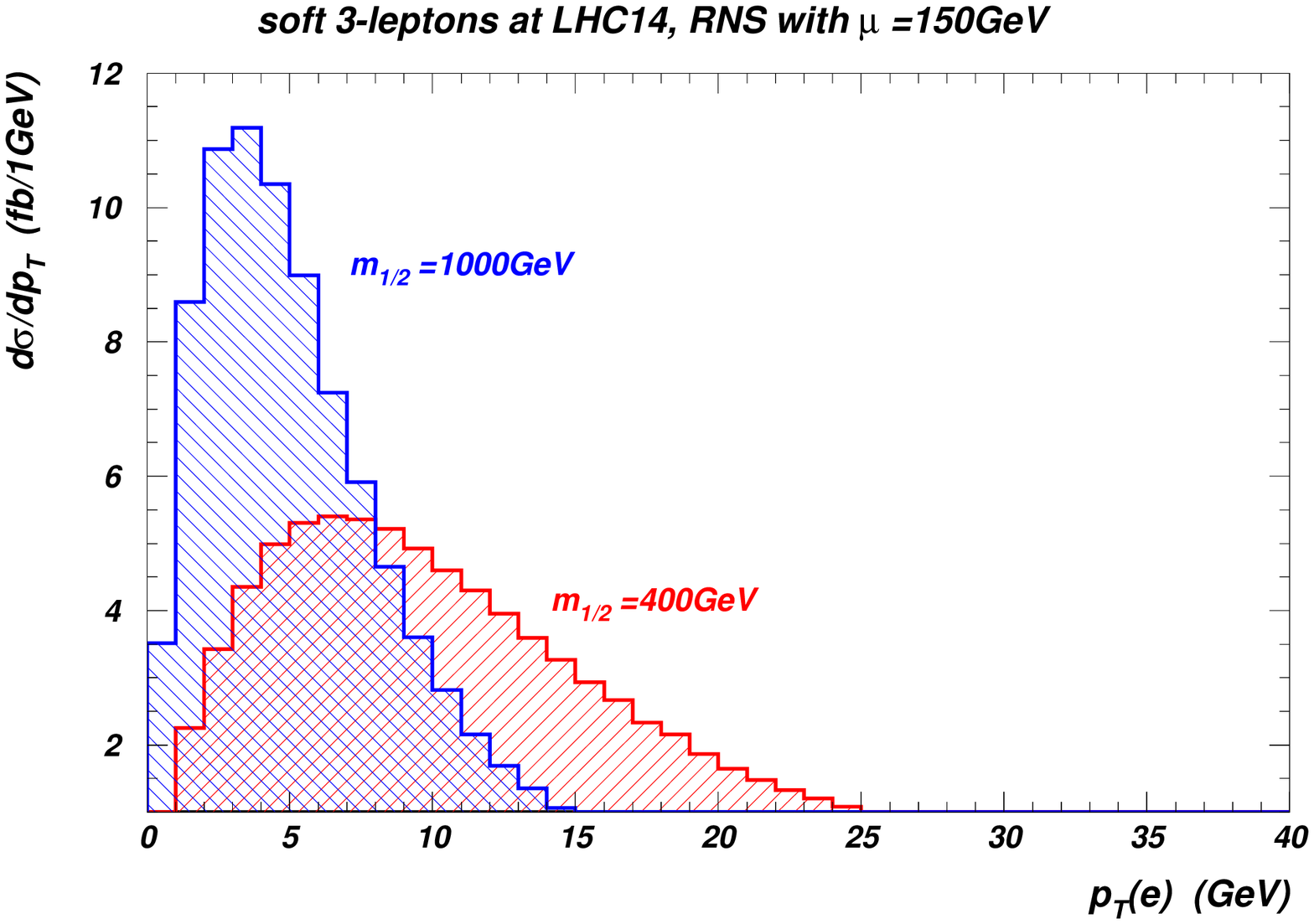}
\caption{$p_T(e)$ distribution for soft tri-leptons from higgsino pairs 
before cuts for two RNS points with 
$m_{1/2}=400$~GeV (red) and 1000~GeV (blue) at LHC14.
}
\label{fig:softe}}

After cut 4, the background exceeds signal by a factor of 10.  The dimuon
invariant mass distribution after cuts is shown in
Fig.~\ref{fig:m(mm)}{\it a}) for $m_{1/2}=400$~GeV ($\tz_2-\tz_1$ mass
gap at 38~GeV), {\it b}) $m_{1/2}=550$~GeV (mass gap at 25~GeV) and {\it
c}) $m_{1/2}=700$~GeV (mass gap at 18~GeV). We see that the {\em shapes}
of the dilepton mass distribution for the signal+background  in frame
{\it a})  differs from that of the background alone. A {\em shape analysis}
using the data at large $m_{\ell\ell}$ to normalize the background may
allow one to claim a signal, given sufficient integrated luminosity, since 
an excess of events should be found in
bins with $m(\mu^+\mu^- )<38$~GeV as compared to higher mass bins where
a theory-experiment match is expected.  For a counting analysis alone, 
invoking a cut
$m(\mu^+\mu^-)<38$~GeV, a $5\sigma$ signal over background (without any
requirement on the $S/B$ ratio) would require
about 700~fb$^{-1}$ of integrated luminosity.  In the other frames with
the smaller mass gap, an excess only appears in the lowest mass bin(s) and
the possibility of extracting a signal appears even more daunting.
\FIGURE[tbh]{
\includegraphics[width=8cm,clip]{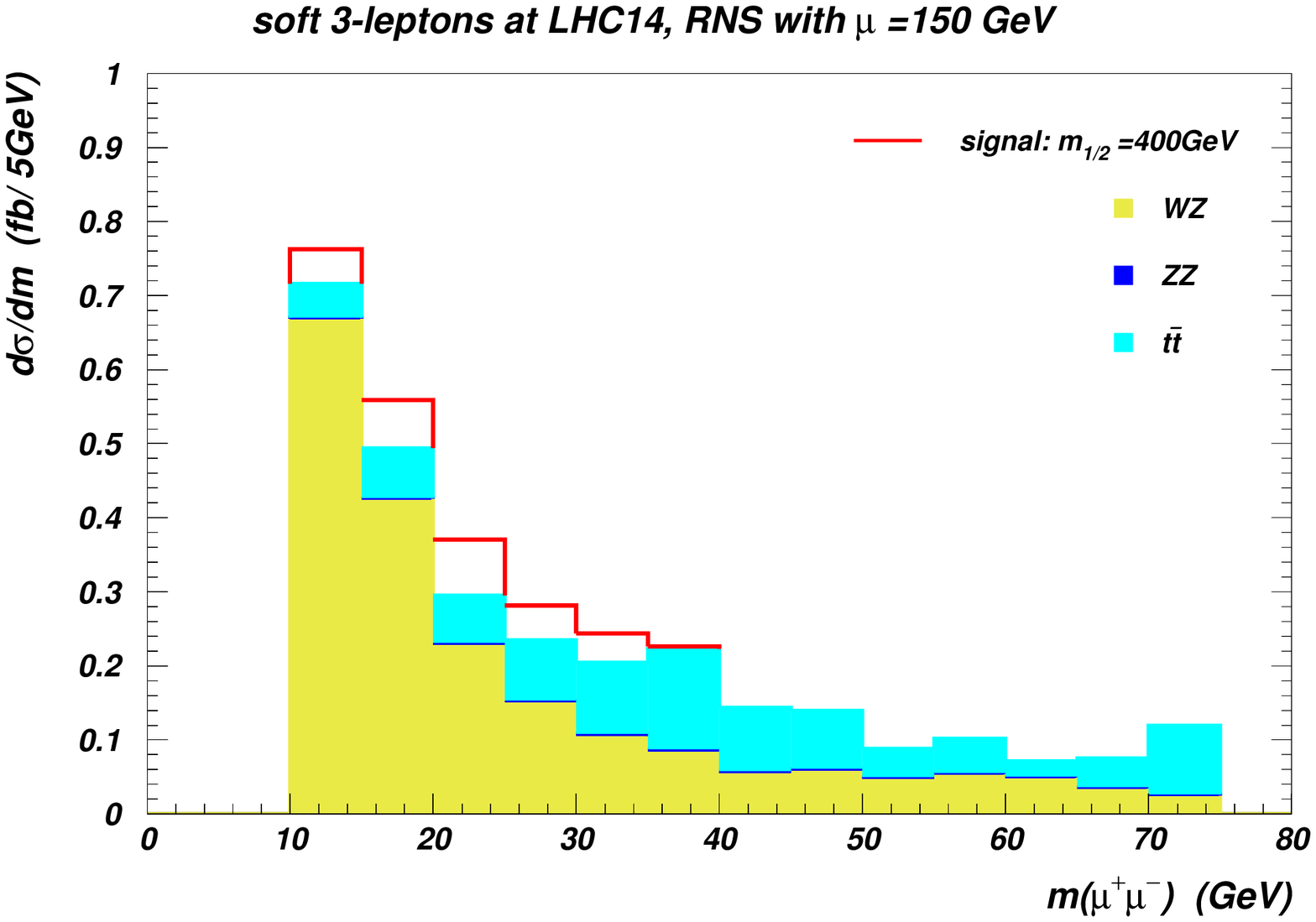}
\includegraphics[width=8cm,clip]{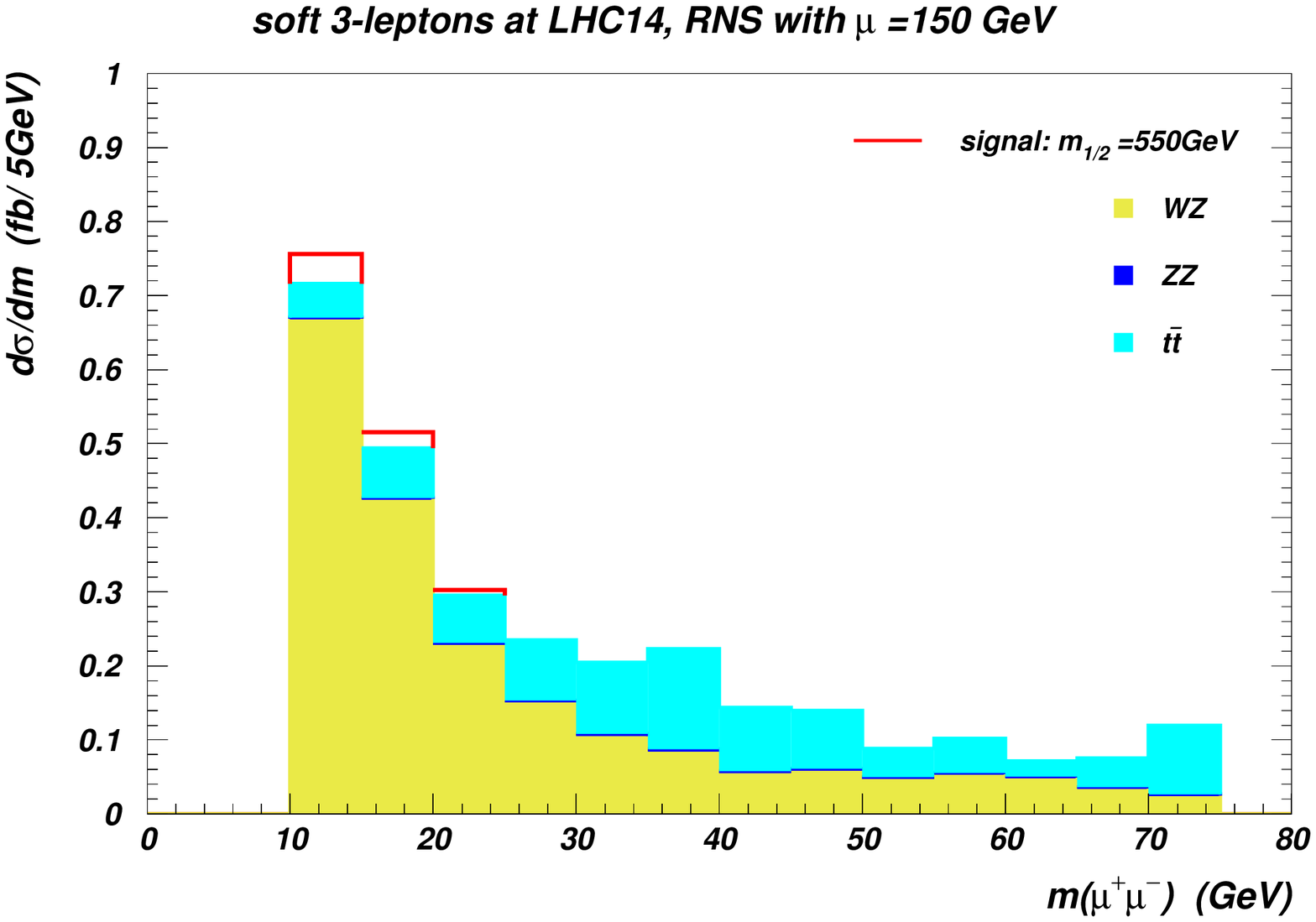}
\includegraphics[width=8cm,clip]{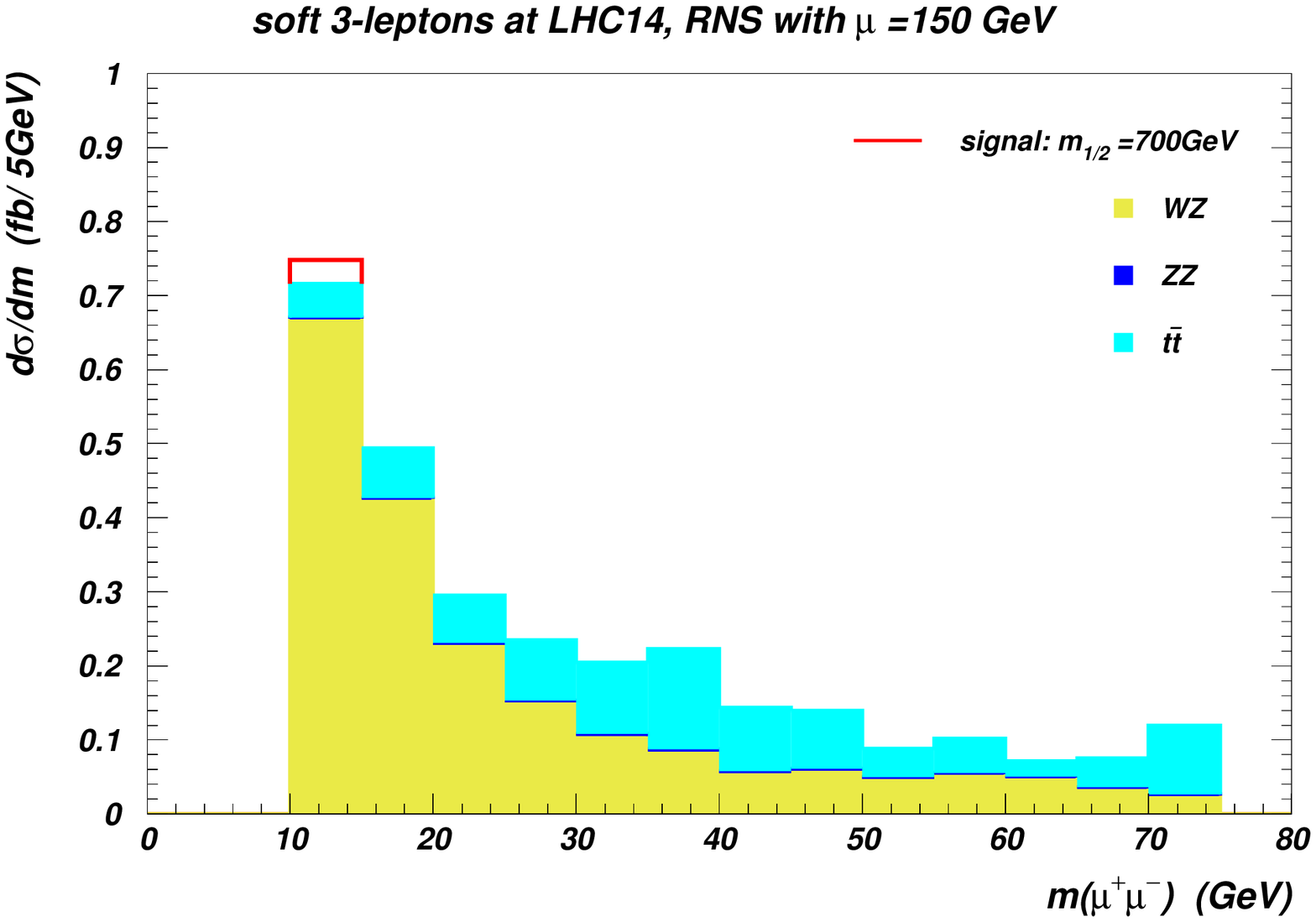}
\caption{The dimuon invariant mass distributions after cuts for $e\mu\mu$ events. 
Open red histograms represent signals from higgsino pair productions for three points 
with $m_{1/2}=400,\ 550$ and 700~GeV along the RNS model line.
}
\label{fig:m(mm)}}
%

\section{Conclusions}
\label{sec:conclude}

Recent results from LHC7 and LHC8 have resulted in heightened concern
for reconciling electroweak naturalness with lack of SUSY signals and
the rather large value of $m_h=125$~GeV.  
We have argued that this reconciliation can occur within the context of
radiatively-driven natural supersymmetry (or RNS) if there is a HS model 
that yields the NUHM2 model with the required correlations 
between parameters  as the GUT-scale effective theory, 
as discussed in Sec.~\ref{sec:intro}.
The RNS model  is characterized by the presence of four light higgsinos
$\tz_{1,2}$ and $\tw_1^\pm$ with masses $\sim 100-300$~GeV, the lower
the better, fine-tuning wise. Top and bottom squarks and gluinos may lie
in the 1-5~TeV range, while first/second generation squarks and sleptons
may lie in the 5-30~TeV range, thus providing at least a partial
decoupling solution to the SUSY flavor and {\it CP} problems.

In this paper, we explored ways to detect RNS at the LHC with
$\sqrt{s}=14$~TeV. To this end, we constructed an RNS model line which
allows variable $m_{1/2}$ while maintaining low electroweak naturalness
$\delew\sim 10-20$ and $m_h\simeq 125$~GeV.  We found that $\tg\tg$
production followed by cascade decays leads to the expected
leptons+jets+$\eslt$ events. These should allow values of $m_{\tg}$ up
to $\sim 1.7$~TeV to be discovered by LHC14 with about
300~fb$^{-1}$ of integrated luminosity. 

A qualitatively new signal, endemic to SUSY models with light
higgsinos, arises mainly from wino pair production $pp\to\tw_2\tz_4\to (W^\pm
\tz_{1,2})+(W^\pm\tw_1^\mp)$ which leads to same sign-diboson production
accompanied by minimal jet activity.  After cuts, the largest background
comes from $t\bar{t}W$ production.  This channel seems to offer the best
reach for RNS for higher integrated luminosity values
$>100$~fb$^{-1}$. The SSdB signal from wino pair production may be confirmed
if the decays of $\tw_2$ and $\tz_4$ yield a final state with $WZ\to 3\ell +\eslt$ 
at high integrated luminosity.  Interestingly, wino pair
production also leads to observable $4\ell+\eslt$ signals
for $m_{1/2} \lesssim 500$~GeV (up to $\sim 650$~GeV at the high-luminosity LHC). 
We also explored signals in the soft $3\ell$ channel arising from direct
higgsino pair production $pp\to\tw_1^\pm\tz_2$. This channel should be
visible over the lower $m_{1/2}$ range, which provide a large enough
$m_{\tz_2}-m_{\tz_1}$ mass gap such that one may avoid the $2\to 4$ process 
$W^*Z^*/W^*\gamma^*$ which contains an obstructing virtual photon contribution at
the lower portion of the $m(\ell^+\ell^-)$ distribution.  Detection will
likely be possible via the analysis of the shape of the dimuon invariant
mass distribution for $e^\pm\mu^+\mu^-$ events where there should be a
distortion due to an excess for $m(\mu^+ \mu^-)<m_{\tz_2}-m_{\tz_1}$.

The final reach situation is summarized in Table~\ref{tab:reach}, where
the $5\sigma$ reach in terms of $m_{\tg}$ is given for various discovery
channels and integrated luminosity values.  While LHC14 can explore RNS
up to $m_{\tg}\sim 2$~TeV for 300~fb$^{-1}$, a large swath of parameter
space with $m_{\tg}\sim 2-5$~TeV seemingly lies beyond LHC14 reach.
\begin{table}
\begin{center}
\begin{tabular}{|l|r|r|r|r|}
\hline
 Int. lum. (fb$^{-1}$) & $\tg\tg$ &  SSdB & $WZ\to 3\ell$ &$4\ell$ \\
\hline
\hline
10   & 1.4 &  --  & -- & --\\
100  & 1.6 &  1.6 & -- & $\sim 1.2$\\
300  & 1.7 &  2.1 & 1.4& $\gtrsim 1.4$ \\
1000 & 1.9 &  2.4 & 1.6& $\gtrsim 1.6$ \\
\hline
\end{tabular}
\caption{Reach of LHC14 for SUSY in terms of gluino mass, $m_{\tg}$ (TeV),  
assuming various integrated luminosity values along the RNS model line.
We present each search channel considered in this paper except soft
$3\ell$.
\label{tab:reach}}
\end{center}
\end{table}

The grand picture is shown in Fig.~\ref{fig:muvsmhf} where we plot the
$\mu\ vs.\ m_{1/2}$ plane of the RNS model, taking the GUT-scale matter
scalar mass parameter $m_0=5$~TeV, $\tan\beta =15$, $A_0=-1.6 m_0$ and
$m_A=1$~TeV.  The green-shaded region has thermal higgsino relic density
$\Omega_{\tilde{h}}h^2<0.12$, which allows for contributions to the dark
matter density from axions\cite{bbl}. The reader may legitimately ask
whether the non-observation of any signal in WIMP direct detection
experiments excludes portions of this region.  In this connection we
should keep in mind that deep in this green region-- {\it i.e.} for
large values of $m_{1/2}$ where the LSP is higgsino-like-- the WIMP {\em
thermal} relic density is strongly suppressed and in fact the dark
matter is expected to be comprised of just 5-10\% Higgsino-like WIMPS along
with the bulk comprising of axions, or something else, over much of parameter
space\cite{bbl}. In such a case, the local WIMP abundance is expected to
be suppressed by a factor 10-15 from the usual assumption, thus allowing
the RNS higgsinos to escape present bounds.  Even with a depleted local
abundance, ton-size noble liquid detectors should be able to test the
entire RNS parameter space with $\Delta_{EW}<100$\cite{bbm}.  The recent
null WIMP search results from the LUX\cite{lux} and XENON100\cite{xenon}
experiments will likely exclude a small region along the edge of the
green region where the thermal neutralino relic density saturates the
observed cold dark matter density. We have not performed a quantitative
study to delineate this region, in part because effects from the sector
responsible for the non-thermal-neutralino component of the dark matter
could significantly modify this analysis. The neutralino contribution to
the cold dark matter could be increased from its thermal expectation due
to late decay of heavy fields to neutralinos, or reduced if new heavy
fields decay to SM particles, leading to entropy-dilution of
neutralinos. In the latter case, axions or some other particles beyond
the MSSM would have to make up the balance.

From Figure~\ref{fig:muvsmhf}, it can be seen that LHC8 has explored
$m_{1/2}< 400$~GeV via search for $\tg\tg$ production. The LHC14 reach
with 300~fb$^{-1}$ for $\tg\tg$ and for same-sign dibosons extends to
$\sim 700-800$~GeV (corresponding to a reach to $m_{\tg}\sim
1.8-2.1$~TeV). The contours of $\delew =30$ extend well beyond the LHC14
reach, to $m_{1/2}\sim 1200$~GeV.
\FIGURE[tbh]{
\includegraphics[width=10cm,clip]{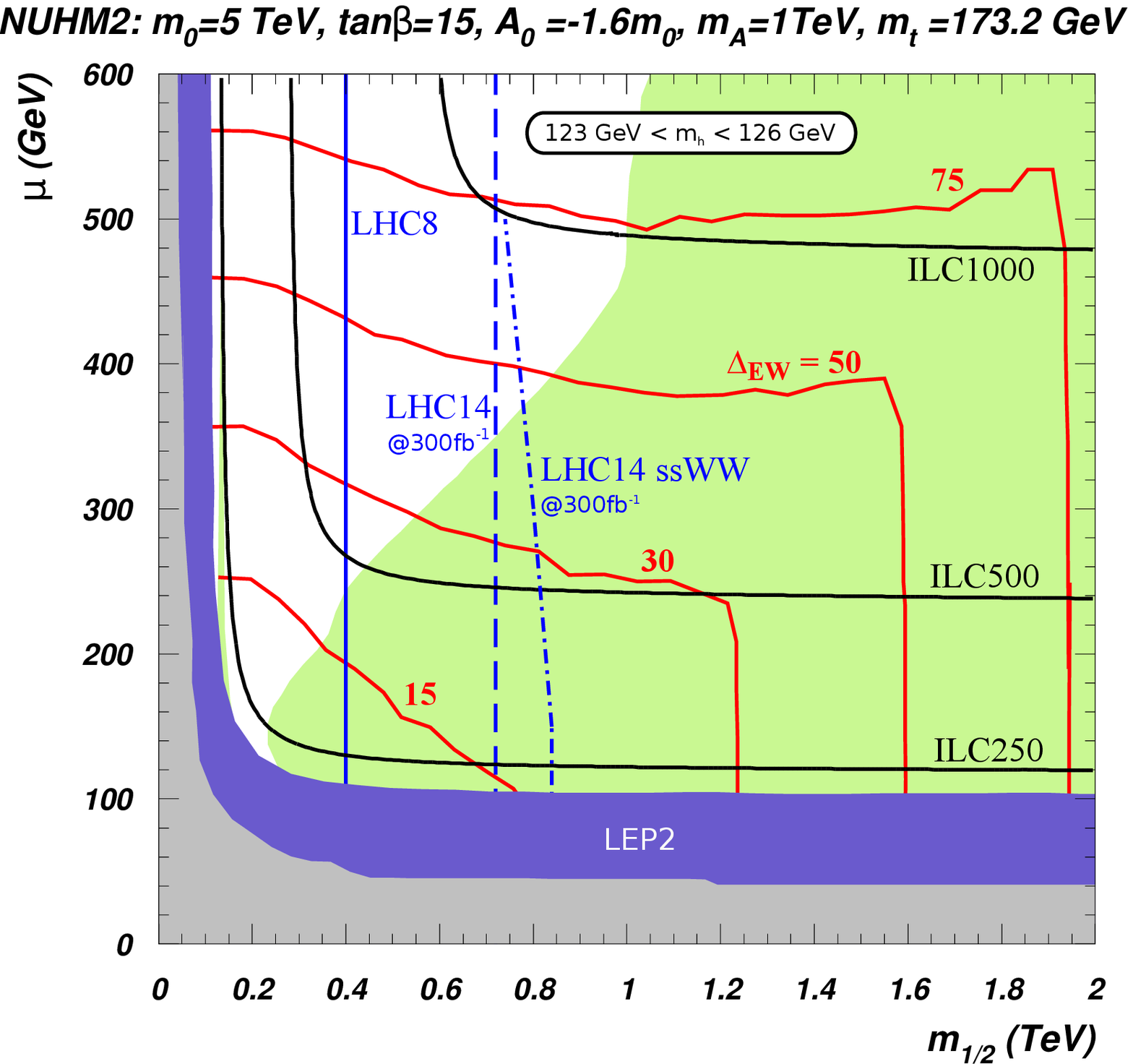}
\caption{Plot of $\delew$ contours (red) in the $m_{1/2}\ vs.\ \mu$
plane of NUHM2 model for $A_0=-1.6 m_0$ and $m_0=5$~TeV and $\tan\beta
=15$.  We also show the region accessible to LHC8 gluino pair searches
(solid blue contour), and the region accessible to LHC14 searches with
300~fb$^{-1}$ of integrated luminosity (dashed and dot-dashed contours).
We also show the reach of various ILC machines for higgsino pair
production (black contours).  The green-shaded region has
$\Omega_{\tz_1}^{std}h^2<0.12$.  The blue (gray) shaded region is
excluded by LEP2 (LEP1) searches for chargino pair production.  To aid
the reader, we note that $m_{\tg}\simeq 2.5 m_{1/2}$.  }
\label{fig:muvsmhf}}
We also show the reach of various $e^+e^-$ collider energy options (ILC
or TLEP) for $\sqrt{s}=250$, 500 and 1000~GeV via the reaction
$e^+e^-\to\tw_1^+\tw_1^-$\cite{list}.  We see that ILC with
$\sqrt{s}\sim 600$~GeV can probe the entire parameter space with $\delew
< 30$, thus either discovering or ruling out the light higgsinos which
are a necessary condition for SUSY naturalness.  Thus, LHC14 plus ILC600
can make a complete search for RNS models which automatically
accommodate electroweak naturalness along with $m_h\simeq 125$~GeV.
These collider signals should also be accompanied by direct and perhaps
indirect dark matter signals from detections of relic
higgsinos\cite{bbm}, which would likely make up only a portion of the
entire dark matter abundance.

\acknowledgments
We thank Baris Altunkaynak for pointing out an error in some 
total cross sections of Fig. 4, 5 in the initial version.
AM would like to thank FTPI at the University of Minnesota for hospitality during final stages of the
project. 
This work was supported in part by grants from the U.S. Department of Energy,  
by Suranaree University of Technology, and by the Higher
Education Research Promotion and National Research
University Project of Thailand, Office of the Higher Education
Commission.
	
%

\end{document}